\documentclass[11pt]{article}
\usepackage[a4paper]{geometry}
\usepackage{jcapmod}
\usepackage{bm}
\usepackage{amsopn}
\usepackage{latexsym}
\usepackage{amsmath}
\usepackage{colortbl}
\usepackage{multirow}
\usepackage{array}
\usepackage{booktabs}
        \usepackage{fancyhdr}
\usepackage{rotating}
\usepackage{graphicx}
        \usepackage[charter]{mathdesign}
        \usepackage{enumerate}

\let\oldsqrt\sqrt
\def\sqrt{\mathpalette\DHLhksqrt}
\def\DHLhksqrt#1#2{%
\setbox0=\hbox{$#1\oldsqrt{#2\,}$}\dimen0=\ht0
\advance\dimen0-0.2\ht0
\setbox2=\hbox{\vrule height\ht0 depth -\dimen0}%
{\box0\lower0.4pt\box2}}

\def\clap#1{\hbox to 0pt{\hss#1\hss}}

\newcommand{\im}{\mathrm{i}}
\newcommand{\EulerGamma}{\gamma_{\mathrm{E}}}

\newcommand{\Mp}{M_{\mathrm{P}}}

\newcommand{\etal}{et al.}

\newcommand{\fNL}{f_{\mathrm{NL}}}

\DeclareMathOperator{\Or}{O}

\DeclareMathOperator{\Ei}{Ei}

\DeclareMathOperator{\RePart}{Re}

\renewcommand{\Re}{\RePart}

\renewcommand{\d}{\mathrm{d}}
\renewcommand{\vec}[1]{\bm{\mathrm{{#1}}}}

\newcommand{\para}[1]{\par\vspace{2mm}\noindent\emph{{#1}}.---}

\newcolumntype{Q}{>{$\displaystyle}l<{$}}
\newcolumntype{q}{>{\columncolor[gray]{0.9}$\displaystyle}l<{$}}
\newcolumntype{R}{>{$\displaystyle}r<{$}}
\newcolumntype{S}{>{$\displaystyle}c<{$}}
\newcolumntype{s}{>{\columncolor[gray]{0.9}$\displaystyle}c<{$}}
\newcolumntype{T}{>{\columncolor[gray]{0.9}}c<{}}

\newsavebox{\tableA}
\newsavebox{\tableB}

\newsavebox{\boxplot}
\newsavebox{\boxplota}

\begin{document}

\begin{flushright}
\begin{small}
DAMTP-2012-13 
\end{small}\end{flushright}

	\title{
	Inflationary signatures of single-field models 
	beyond slow-roll}

	\author{Raquel H. Ribeiro}

	\affiliation{Department of Applied Mathematics and Theoretical Physics\\
	Centre for Mathematical Sciences, University of Cambridge \\
	Wilberforce Road \\
	Cambridge CB3 0WA, United Kingdom} 

	\emailAdd{R.Ribeiro@damtp.cam.ac.uk}

	\abstract{If the expansion of the early Universe was not close to de Sitter, 
	the statistical imprints of the primordial density perturbation
	on the cosmic microwave background can be quite different from those
	derived in slow-roll inflation.
	In this paper we study the inflationary signatures of 
	all single-field models  which are
	free of ghost-like instabilities.
	We allow for a rapid change of the Hubble
	parameter and the speed of sound of scalar fluctuations, in a 
	way that is compatible with a nearly scale-invariant spectrum of perturbations, 
	as supported by current cosmological observations.
	Our results rely on the scale-invariant approximation, which is 
	different from the standard slow-roll approximation.
	We obtain the propagator of scalar fluctuations and compute the bispectrum, 
	keeping next-order corrections proportional
	to the deviation of the 
	spectral index from unity. 
	These theories offer an explicit example where the shape 
	and scale-dependences of the bispectrum are highly non-trivial 
	whenever slow-roll is not a good approximation.}


\vspace*{2cm}

	\maketitle

	\section{Introduction}
	\label{sec:introduction}

In recent years our understanding of the early Universe has become much clearer
as we gain access to precise measurements of 
the Cosmic Microwave Background Radiation (CMBR) 
\cite{Komatsu:2010fb, Larson:2010gs, Jarosik:2010iu}.
It is widely believed that in the early Universe a period of inflationary expansion took place
with cosmological scales being pushed outside the observable horizon \cite{Guth:2005zr}. 
With this simple idea the Universe becomes very smooth, the
number density in relics is diluted away \cite{Guth:1980zm}, and the 
standard cosmological problems are ameliorated. 
At the same time, 
distinctive signatures are imprinted in the 
CMBR anisotropies which can be observed by 
high-resolution surveyors, such as the \textit{Planck} satellite.

From CMB 
and Large Scale Structure (LSS) data 
\cite{Peiris:2009wp} we know that the spectrum of scalar perturbations
is nearly scale-invariant, with the power being 
only slightly stronger at larger scales. Indeed, 
according to the most recent WMAP$7$ data analysis, 
the spectral index of density perturbations evaluated at 
$k=0.002 \ \textrm{Mpc}^{-1}$ is $n_s=0.969 \pm 0.012$ \cite{Larson:2010gs}.
Unfortunately,
a scale-invariant spectrum is a model-independent feature of inflation. 
To look for distinctive 
signatures of different models it is necessary to study 
higher-point statistics of the primordial
curvature perturbation, 
known as non-gaussianities \cite{Komatsu:2009kd}---they result from the 
inflaton self-interactions and couplings to the gravity sector.

Amongst the most important of these statistics 
and potentially easiest to observe is
the bispectrum or, in quantum field theory language, 
the three-point correlation function.
The first studies of the bispectrum started with Maldacena's pioneering work
devoted to canonical single field-models \cite{Maldacena:2002vr}, later generalized
by Seery \& Lidsey \cite{Seery:2005wm} and 
Chen \etal \, \cite{Chen:2006nt} for theories
with a varying speed of sound for scalar 
fluctuations.\footnote{For simplicity we refer
						to the phase speed of perturbations
						\cite{Christopherson:2008ry}
						as the speed of sound.}
The latter class of models became known 
as $P(X,\phi)$ theories, because the Lagrangian is an arbitrary 
function, $P$, of the field profile, $\phi$, and its first derivatives
through $X\equiv - (\nabla\phi)^2$.
These models have the property of allowing for large 
non-gaussianities whenever the speed of sound for perturbations
is small.
On the other hand,
effective field theories described by canonical kinetic 
terms and higher powers of single-derivative operators 
(suppressed by powers of some high energy physics scale), 
predict observationally small non-gaussianities, 
if the effective field theory description is to remain valid \cite{Creminelli:2003iq}.
All the calculations mentioned above used the slow-roll approximation.
For more recent works see, for example, 
Refs. \cite{Chen:2009zp, Chen:2009we, Creminelli:2010qf, Bartolo:2010bj, 
Burrage:2010cu, Kamada:2010qe, Baumann:2011su, 
RenauxPetel:2011dv, RenauxPetel:2011uk, Ribeiro:2011ax}.

To understand better the microphysics processes which operated in the early Universe
one needs to fully understand these non-gaussian signatures, since ultimately
they have the power to eliminate models against observations.
With \textit{Planck}'s data
soon to become available our best hope to achieve this
 lies within a non-zero measurement of the bispectrum, 
 which will be the object of interest in this paper.

The simplest inflationary models are those where 
the scalar field theory contains only one
scalar degree of freedom responsible for sourcing inflation and  
seeding perturbations: the inflaton.\footnote{There exist more complicated
										models with multiple fields, for example  
										sufficiently heavy so that they can be integrated out, 
										resulting in
										an effective field theory description of
										the inflationary dynamics---see, for example,
										Refs. \cite{Achucarro:2012sm, Avgoustidis:2012yc}.}
In spite of their simplicity, these models accommodate quite a 
remarkable variety of interesting scenarios: from 
models with canonical 
kinetic terms, to Dirac--Born--Infeld 
(DBI)
inflation \cite{Silverstein:2003hf, Alishahiha:2004eh}  and
galileon inflation \cite{Burrage:2010cu}.
These last two models
are the only known radiatively stable theories
with higher order derivatives, because their action 
is protected by special symmetries.
On the other hand, a number of authors have rather
focused on the much milder requirement that 
the Lagrangian preserved unitarity,
resulting in a sensible quantum field theory
\cite{Deffayet:2010qz, Kobayashi:2010cm, Mizuno:2010ag, Kobayashi:2011nu, 
Kobayashi:2011pc, Gao:2011qe}.
In these theories the galilean symmetry, 
which was the main motivation behind galileon theories, is lost
because it can no longer be realised in 
a generic spacetime \cite{Deffayet:2009wt, Hui:2012qt}.
The Lagrangian operators in the action give rise to equations of motion 
for the scalar field which are at most second-order in derivatives.
Recently these general theories were shown to be equivalent to
 Horndeski models \cite{Kobayashi:2011nu, Gao:2011qe}. First derived 
over thirty years ago \cite{Horndeski:1974},
Horndeski theories are described by the most generic action 
involving one scalar field and do not contain ghost-like instabilities, 
hence encapsulating all the single-field models of potential interest.

Until recently the traditional methodology to compute the bispectrum 
started from the action for 
the comoving curvature perturbation, $\zeta$ \cite{Bardeen:1983qw, Mukhanov:1990me}, 
for each model independently.
This perturbation is the appropriate quantity to study since it is conserved on 
super-horizon scales, if isocurvature modes are absent.
However, new developments have simplified this procedure. 
First, Gao \& Steer \cite{Gao:2011qe} and 
De Felice \& Tsujikawa \cite{DeFelice:2011uc} 
obtained the universal action for perturbations in stable single-field models
involving what we call Horndeski operators
(these are the operators appearing in the cubic action for $\zeta$). 
The only model dependent features in the action reside in different 
coefficients of each operator.
Second, 
it was shown in Ref. \cite{RenauxPetel:2011sb} that the cubic action for $\zeta$ has a 
minimal representation in terms of only \textit{five} of these Horndeski operators, 
initially derived in Ref. \cite{Burrage:2011hd}.

With these latest developments we arrive at a universal methodology 
to compute the bispectrum
of all single-field models, 
with the cubic action for $\zeta$ being always of the form
	\begin{equation}
		S^{(3)} = 
		\int \d^3 x \, \d\tau \; \Big\{ 
			a \Lambda_1 \dot{\zeta}^3 +a^2 \Lambda_2 \zeta \dot{\zeta}^2
			+ a^2 \Lambda_3 \zeta (\partial \zeta)^2
			+ a^2 \Lambda_4 \dot{\zeta} \partial_i \zeta
				\partial^i (\partial^{-2}\dot{\zeta})
			+ a^2 \Lambda_5 \partial^2 \zeta
				(\partial_i \partial^{-2}\dot{\zeta})^2	
		\Big\} \ ,
		\label{eq:zeta-action3}
	\end{equation}
where $a$ is the scale factor, dotted quantities are differentiated with 
respect to conformal (not cosmological) time
and $\partial$ denotes a spatial partial derivative.
This form of the action was first derived in Ref. \cite{Burrage:2011hd},
although with different coefficients $\Lambda_i$.\footnote{This action had appeared first in Refs. 
							\cite{Seery:2005wm, Chen:2006nt} specialized for $P(X,\phi)$ models.
							There, the action involved a larger number of cubic operators 
							and also a field redefinition. The equivalence of the 
							actions therein and in Eq. \eqref{eq:zeta-action3} was explained
							in Ref. \cite{Burrage:2011hd}.}
As we mentioned before, 
the model-dependent imprints will be encoded in each of the 
\textit{five} coefficients $\Lambda_i$ of 
the Horndeski operators. There is a priori no 
hierarchy between these coefficients, although specialization 
to different models can impose specific ratios between the $\Lambda_i$ 
(as in DBI inflation).
The action above will be our starting point in computing the bispectrum
of Horndeski theories.

Many authors have focused on the study of the action 
\eqref{eq:zeta-action3} in the
slow-roll regime, in which the inflationary expansion 
is quasi-de Sitter, whilst
allowing for small variation of the sound speed of perturbations. 
In this approximation, $\varepsilon\equiv -\d \ln H/\d N$, 
$\eta\equiv \d \ln \varepsilon /\d N$
and $s\equiv \d \ln c_s/\d N $ all obey 
$\varepsilon, |\eta|, |s| \ll 1$, 
where $N$ stands for the number of e-folds.
Examples of such works include
Refs. \cite{Seery:2005wm, Chen:2006nt, Burrage:2010cu, Burrage:2011hd, Ribeiro:2011ax}.
But, what if inflation was \textit{not} almost de Sitter, and the parameters 
above cannot be treated perturbatively?
Khoury \& Piazza \cite{Khoury:2008wj} showed that this scenario was still
compatible with a scale-invariant spectrum of perturbations,
in fair agreement with observations \cite{Larson:2010gs}, 
provided the relation $s=-2 \varepsilon$ was satisfied.\footnote{These authors have also considered 
													solutions within the ekpyrotic 
													mechanism (see Ref. \cite{Lehners:2008vx} 
													for a review, and 
													also Refs. \cite{Khoury:2010gw, Baumann:2011dt}). 
													In this paper however, we 
													focus on the inflationary scenario.} 
We know that both $\varepsilon$
and |$\eta$| must be small to allow for a successful 
period of inflation. 
But, \textit{how small} are they required to be? 
If $\varepsilon$ is not necessarily much smaller than $1$, 
then a calculation beyond slow-roll is technically required,
especially
to fit in the current era of precision Cosmology.
Previous works which have attempted to study correlations
beyond the slow-roll regime include corrections to the 
power spectrum in canonical models studied by 
Steward \& Lyth \cite{Stewart:1993bc}; also, the work by  
Gong \& Stewart \cite{Gong:2001he} 
(and Ref. \cite{Choe:2004zg}) 
who applied the Green's function method to obtain the propagator of scalar fluctuations
to next-next-order in slow-roll---these corrections 
are second order higher than the leading-order results in
the slow-roll approximation invoked in these calculations 
(see below for details on this terminology).
Their work was later generalized by Wei \etal\, \cite{Wei:2004xx}  who considered models
with a varying sound speed of perturbations.
Additionally Bartolo \etal \, \cite{Bartolo:2010im} have calculated
the power spectrum beyond leading order in the context of
effective field theories of inflation \cite{Cheung:2007st, Weinberg:2008hq}.

In this paper we will be interested in models where the spectrum is almost scale-invariant, 
so that the condition found by Khoury \& Piazza is mildly broken and becomes
$s=-2\varepsilon+\delta$. 
This requires working to \textit{all} orders in $\varepsilon$ and $s$, 
but \textit{perturbatively} in $\delta$, 
and time variations of $\varepsilon$ and $s$. This last assumption
ensures that the conditions $\varepsilon, |s|<1$ are preserved for a sufficiently
large number of e-folds, which is a primary requisite for 
 inflation to last at least $60$ e-folds.\footnote{This work explicitly excludes models with 
								features in the potentials, which can trigger the 
								slow-roll parameters to temporarily grow 
								during inflation \cite{Chen:2010xka}.
										}
In this sense, this is a calculation \textit{beyond} the
slow-roll approximation, since it requires a resummation technique
applied to the non-perturbative parameters $\varepsilon$ and $s$. 
We refer to it as \textit{scale-invariant 
approximation}. 
Starting from the action \eqref{eq:zeta-action3}, the calculation 
is immediately applicable to all Horndeski models.
For this reason we will work with arbitrary 
interaction coefficients $\Lambda_i$---assignments to these coefficients
will correspond to specific models.

We organise the 
calculation in increasing powers in the hierarchy of slow-variation parameters, 
focusing on leading and next-order contributions. 
Leading order results involve the least power of perturbative 
parameters, and correspond to results obtained assuming 
a perfectly scale-invariant spectrum of perturbations. 
Next-order corrections contain small deviations 
from this which are parametrized by $n_s-1$, and therefore involve 
terms with one extra power in 
the perturbative parameters.
This organisation scheme follows the one used throughout 
Refs. \cite{Burrage:2010cu, Burrage:2011hd, Chen:2006nt}
which nevertheless applied the \textit{slow-roll approximation}.
On the other hand, our calculation relies on the 
\textit{scale-invariant approximation} and generalizes their results
 up to next-order in scale-invariance.
We explicitly obtain the scale-dependence of the bispectrum, 
which appears more intricate than the one found with the assumptions 
of slow-roll. In particular, the bispectrum exhibits a 
strong, power-law scale-dependence, accompanied by the traditional
weak, logarithmic-dependence.

\para{Outline}
This paper is organised as follows.
In \S\ref{sec:dynamics} we obtain the dynamical behaviour
of the scale factor and the other relevant background quantities
in the scale-invariant approximation.
The dynamical behaviour is \textit{exact} in $\varepsilon$ and $s$, 
and \textit{perturbative} in the time variation of these parameters, $\delta$ and $n_s-1$.
In \S \ref{sec:power spectrum} we obtain the power spectrum for scalar fluctuations in this 
scale-invariant approximation,
and we derive in \S\ref{sec:propagator} formulae for
 the elementary wavefunctions,
from which the scalar propagator is built,
to next-order in scale-invariance.
We use these results to 
compute the bispectrum of perturbations in \S\ref{sec:bispectrum}
and comment on the differences when using 
the slow-roll approximation. 
We conclude in \S\ref{sec:conclusions}.
The appendices collect derivations of 
formulae used throughout the text and detailed expressions 
of recurring integrals necessary to produce closed form bispectra.
\para{Notation}	
	We choose natural units in which $c = \hbar = 1$
	and we take the reduced Planck mass, $\Mp = (8 \pi G)^{-1/2}$, to be unity.
	The metric signature is $(-,+,+,+)$,
	and Latin letters $\{ i, j, \ldots \}$ denote purely spatial indices, which are also 
	occasionally used 
	to refer to each Horndeski operator.
Derivatives with respect to conformal time, $\tau=\int_{\infty}^{t}{dt/a(t)}$, 
will be denoted 
by dots, and not primes.

\section{Background evolution beyond exact scale-invariance}
\label{sec:dynamics}
We are interested in studying the inflationary signatures of
 nearly scale-invariant models which fall under
the class of Horndeski theories. As argued in Ref. \cite{Khoury:2008wj},
a scale-invariant spectrum of perturbations can be accommodated 
by a background where both the Hubble parameter and the speed of sound of perturbations
vary significantly in time, 
provided they obey the \textit{exact} relation $s=-2 \varepsilon$. 

We start with an arbitrary Horndeski theory for a homogeneous
scalar field, on which small, inhomogeneous perturbations, $\delta \phi$, develop. 
All we require is the spectrum of perturbations to be close to scale-invariant.
The comoving curvature perturbation is related to these perturbations
via $\zeta=aH\delta \phi /\dot{\phi}$. 
In single-field models, one often computes the correlation functions
in the comoving or $\zeta$-gauge rather than in the 
uniform-density or $\delta\phi$-gauge. This has an 
important advantage: 
the correlation functions of $\zeta$ are time-independent
whereas the ones for $\delta \phi$ are not. 
For this reason it is more convenient to work with 
the action for perturbations written as Eq. \eqref{eq:zeta-action3}.
We will come back to this point shortly.

For all inflation models involving one single clock in the Universe, 
the quadratic action for $\zeta$ can be written as 
\begin{equation}
S^{(2)}=\dfrac{1}{2}\int \d^3 x \d\tau \, a^2 \, z\,
\bigg\{ \dot{\zeta}^2 -c_s^2 \big(\partial \zeta  \big)^2\bigg\} \ \ ,
\label{eq:quadratic action1}
\end{equation}
where $z$ is required to be a well defined, differentiable function of
the background dynamics and Lagrangian parameters,
but it is otherwise arbitrary.\footnote{For $P(X,\phi)$ models, $z=\varepsilon/c_s^2$
									 \cite{Burrage:2011hd},
									but the same need not be true for other models 
									like galileon inflation theories
									\cite{Burrage:2010cu}, for example.
									We will come back to this point in \S\ref{sec:propagator}.
									We note that $z$ here is not the usual $z$ 
									for $k$-inflation.}
The time evolution of $z(y)$ will be parametrized by $w\equiv \d \ln z/\d N$.
For consistency, we also work to all orders in $w$.
For an arbitrary $w$, Khoury \& Piazza's relation between $\varepsilon$
and $s$ required for scale-invariance, 
is generalized to\footnote{We refer the reader to Eq. \eqref{eq:qsi_relation}
										in appendix \ref{app:ns} for details
										on 
										the derivation of this formula.}
\begin{equation}
3s=-2\varepsilon-w\ \ . 	
\label{eq:kpsimple}
\end{equation}
This formula reduces to $s=-2\varepsilon$ for $P(X,\phi)$ models
and constant $\varepsilon$.				

As discussed in appendix \ref{appendix:y},					
it is more convenient to study the dynamics of background 
quantities as a function of $y$, satisfying $dy=c_s d\tau$, 
rather than conformal time, 
$\tau$.
In this new time-coordinate, action \eqref{eq:quadratic action1} then becomes
\begin{equation}
S^{(2)}=\dfrac{1}{2}\int \d^3 x \d y \, q^2\,
\Big\{  \zeta'^2 -\big(\partial \zeta  \big)^2\Big\}\ \ ,
\label{eq:quadratic action2}
\end{equation}
with $q=a\sqrt{z\,c_s}$ and where the prime denotes derivative
with respect to $y$, and not conformal time. 
Introducing the canonically normalised field 
$v_k=q \zeta_k$, where the subscript $k$ specifies
the Fourier mode, and
changing variables to $v_k=A_k \sqrt{-k y}$, we find that the 
modes $A_k$ obey
a Bessel equation of the form
\begin{equation*}
A_k''+\dfrac{1}{y} A_k'+
A_k \, \bigg\{k^2- \bigg(\dfrac{q''}{q}+\dfrac{1}{4y^2}\bigg)   \bigg\} =0 \ ~\ .
\end{equation*}
The solution to this equation is a linear combination of Hankel functions, 
$ H^{(1)}_\nu (-ky)$ and $ H^{(2)}_\nu (-ky)$, 
with order 
\begin{equation*}
\nu=\dfrac{3}{2}-\dfrac{n_s-1}{2}\ \ .
\end{equation*}
For a perfectly scale-invariant spectrum of perturbations $\nu_0=3/2$
and the spectral index is unity. This results in the power being 
constant in all scales. We refer the reader to appendix \ref{app:ns} for details of
how the spectral index relates to the other background quantities, although the 
precise formula will be unimportant in what follows, 
as we shall see.

Agreement with the appropriate normalization of the propagator gives the 
evolution of the elementary wavefunction as follows
\begin{equation}
\zeta_k=\sqrt{\dfrac{\pi}{8k}} \, \dfrac{\sqrt{-ky}}{a\sqrt{z c_s}}\ H^{(1)}_\nu (-k y)\ \ .
\label{eq:zeta_k}
\end{equation}
Finally, the power spectrum of perturbations can be obtained by
evaluating the scalar propagator at equal times, yielding
\begin{equation}
\langle \zeta (\vec{k_1}) \zeta (\vec{k_2})\rangle =(2\pi)^3\, 
\delta^{(3)} (\vec{k_1}+\vec{k_2}) \dfrac{\pi}{8} 
\dfrac{(-y)}{a^2(y)z(y) c_s(y)} \, \left|H^{(1)}_\nu (-ky)  \right|^2\ \ .
\label{eq:2pf1}
\end{equation}
To further simplify this expression and consolidate 
the time evolution, we will require the evolution of the
scale factor $a$ in $y$ coordinates. In order to do so, we 
specify the parameters in the slow-variation catalogue, 
which will allow us to perform a uniform expansion
up to next-order terms in the scale-invariant approximation, 
as advertised in the introduction.

\vspace*{0,3cm}

\para{Slow-variation catalogue}
To obtain the dynamical evolution of the scale factor in a background
where 
$\varepsilon$ and $s$ are smaller than $1$ 
(but not necessarily much smaller than $1$),
we introduce the following slow-variation parameters:
\begin{equation}
\eta \equiv \dfrac{\d\ln\varepsilon}{\d N} \ \ \textrm{and} \  \ 
t\equiv \dfrac{\d\ln s}{\d N}\ \ .
\end{equation}
We assume these parameters satisfy $|\eta|, |t|\ll 1$.
Our results rely on the scale-invariant expansion, and are 
organised in leading order contributions, and terms contributing at next-order only.
Leading order results are the lowest 
order terms which are compatible with a perfectly scale-invariant 
power spectrum of perturbations, whereas next-order terms are corrections parametrizing the deviation from 
$n_s-1=0$.
This organizational scheme is motivated by the cosmological
data that suggests that $n_s-1$ is very small, and can 
therefore
be treated perturbatively.

To determine the behaviour of the scale factor,
we start by integrating $ c_s d\tau$
to compute $y$. Working \textit{perturbatively} in $\eta$ and $t$,
but to \textit{all} orders in $\varepsilon$ and $s$, we find
\begin{equation*}
y=-\dfrac{c_s}{aH} \sum_{m=0}^{+\infty}\Big\{(\varepsilon+s)^m +
 (\varepsilon+s)^{m-1} \sum_{k=0}^m{k\big(ts +\varepsilon \eta  \big)} \Big\}\ \ .
\end{equation*}
These sums converge and give the dynamical behaviour of $a$ in 
$y$-time 
\begin{equation}
a(y)=-\dfrac{c_s}{Hy\,(1-\varepsilon-s)}\, 
\bigg\{1+\dfrac{\varepsilon \eta +t s}{(1-\varepsilon -s)^2} \bigg\} \ \ .
\label{eq:a}
\end{equation}
This result was first obtained by Khoury \& Piazza in Ref. \cite{Khoury:2008wj}.
Using Eq. \eqref{eq:a} we can further simplify the two-point correlator
\eqref{eq:2pf1} in 
$y$-space, which becomes
\begin{equation}
\langle \zeta (\vec{k_1}) \zeta (\vec{k_2})\rangle =(2\pi)^3\, 
\delta^{(3)} (\vec{k_1}+\vec{k_2}) \dfrac{\pi}{8} 
\dfrac{H^2(y) (-y)^3  (1-\varepsilon-s)^2 }{z(y)\, c_s^3(y)} \,
\bigg\{1-2\dfrac{\varepsilon \eta +t s}{(1-\varepsilon -s)^2} \bigg\} 
 \left|H^{(1)}_\nu (-ky)  \right|^2\ \ .
\label{eq:2pf2}
\end{equation}

In a generic theory, the two-point correlator will evolve in time. 
However, for a single-field model, by virtue of conservation of 
$\zeta$ on super-horizon scales
 \cite{Lyth:1984gv, Wands:2000dp}, 
it follows that the power spectrum 
will rapidly converge to its 
asymptotic value.\footnote{See Ref. \cite{Nalson:2011gc} for comments
							on the number of e-folds necessary for such 
							asymptotic behaviour to be reached
							after horizon crossing. There, the authors study 
							a two-field inflation model, but the same conclusions
							 apply in single-field models.}
Therefore, we only need to focus our calculation a few e-folds after horizon crossing
for a given scale, $k_\star$. This has become the standard approach
when calculating the asymptotics of the power spectrum \cite{Burrage:2011hd, Chen:2006nt}.
This remains true even in multi-field inflation, 
where this procedure sets the appropriate 
initial conditions for evolving correlation functions
after horizon crossing, using techniques such as $\delta N$
\cite{Lyth:2005fi}, 
transport equations \cite{Mulryne:2009kh, Mulryne:2010rp} 
or transfer matrices \cite{Wands:2002bn} (see, for example,
Refs. \cite{GrootNibbelink:2001qt, Lalak:2007vi, Peterson:2010np, 
Peterson:2010mv, Avgoustidis:2011em, Peterson:2011yt}). 
The advantages of applying this approach are twofold: 
on one hand, at this point in the evolution the elementary wavefunction
is completely characterized by the growing mode, since the decaying mode has 
become negligible; on the other hand, the power spectrum 
evaluated a few e-folds after horizon crossing
is already the asymptotic value, at late times.
Said differently, at this evaluation time, $\zeta$ has already classicalized \cite{Lyth:1984gv, Lyth:2006qz}.

With this methodology in mind, and working perturbatively in $\eta$ and $t$,
we find that to next-order in scale-invariance
	\begin{subequations}
		\begin{align}
			\varepsilon 
			& \approx \varepsilon_{\star} 
			\Big\{1-\dfrac{\eta_\star}{1-\varepsilon_\star-s_\star} \ln(-k_\star y) \Big\}
			\ \ \textrm{and}\\ 
			s 
			& \approx s_{\star} 
			\Big\{1-\dfrac{t_\star}{1-\varepsilon_\star-s_\star} \ln(-k_\star y) \Big\}		\ \ .
		\end{align}
					\label{eq:sr1}
	\end{subequations}
In the remainder of this paper we employ the same notation as Refs.
\cite{Burrage:2010cu, Burrage:2011hd} where 
starred quantities are to be evaluated when some reference scale, $k_\star$, 
has exited the $y$-horizon, that is, when $k_\star y_\star=-1$. 
It is important to leave this 
scale arbitrary since this will produce the scale-dependence
of the correlation functions. The appearance of the logarithmic 
term corresponds precisely to the number of 
e-folds, $N_\star$, in $y$-coordinates, 
which have elapsed since the scale $k_\star$
has exited the horizon.\footnote{The notation for the number of e-folds
								as a measure of expansion
								was initially defined by Sasaki \& Tanaka
								in Ref. \cite{Sasaki:1998ug}.}
As argued above, the expansion in Eqs. \eqref{eq:sr1} is only valid up to a 
few e-folds outside the horizon, when one can trust the expansion in 
Taylor series to first order.
We note that nearly scale-invariance demands $\eta_\star=t_\star$, 
up to next-next-order corrections, 
which are beyond the scope of this paper.\footnote{Next-next-order corrections 
									are two orders higher than leading order results.
									We expect next-next-order terms to be smaller 
									than next-order results generically by the same amount 
									that next-order
									terms are corrections to the leading-order calculation.}

The individual dynamics of $c_s(y)$ and $H(y)$ is derived in 
Eqs. \eqref{eq:cs} and \eqref{eq:h} of  
appendix \ref{app:propagator}. These formulae can be used to 
replace for the evolution of the scale factor in
Eq. \eqref{eq:a}. We find:
\begin{equation}
a(y)=\dfrac{c_{s\star}(-k_\star y)^{-\frac{s_\star+\varepsilon_\star}{1-\varepsilon_\star-s_\star}}}{H_\star (1-\varepsilon_\star-s_\star) (-y)} 
\bigg\{ 
1+\beta_\star
-\bigg[ \dfrac{\beta_\star}{1-\varepsilon_\star - s_\star}\bigg] \ln(-k_\star y)+
\bigg[\dfrac{\alpha_\star (\varepsilon_\star +s_\star)}{2(1-\varepsilon_\star -s_\star)}\bigg]\big(\ln(-k_\star y)\big)^2 
\bigg\} \ \ ,
\label{eq:scale_factor}
\end{equation}
where the abbreviated notation for the  
next-order parameters, $\alpha$ and $\beta$,
is defined in Eqs. \eqref{eq:alphabeta}.
Despite its complicated structure, Eq. \eqref{eq:scale_factor} 
correctly reproduces the results 
of Refs. \cite{Khoury:2008wj, Burrage:2011hd} 
when we take the limit of \textit{exact} scale-invariance, 
for which $\eta_\star=t_\star=0$, resulting in vanishing 
$\alpha_\star$ and $\beta_\star$ [cf. Eqs. \eqref{eq:alphabeta}].
The extra terms in Eq. \eqref{eq:scale_factor} are precisely
the corrections to a pure, dominant power-law evolution.

\subsection{The scalar power spectrum}
\label{sec:power spectrum}

As mentioned above, the power spectrum $P(k)$ defined by
\begin{equation}
\langle \zeta (\vec{k_1}) \zeta (\vec{k_2})\rangle =(2\pi)^3\, 
\delta^{(3)} (\vec{k_1}+\vec{k_2})  P(k)\ \ ,
\end{equation}
where $k=|\vec{k}|=|\vec{k}_1|=|\vec{k}_2|$,
will be in general time-dependent. Taking the super-horizon limit, $|ky|\rightarrow 0$,
we find that the dominant contribution arising from the Hankel function
in Eq. \eqref{eq:2pf2} is given by
\begin{equation*}
\left|H^{(1)}_\nu (-k y)   \right|^2 \longrightarrow -\dfrac{2}{\pi k^3 y^3} 
\Big\{1+ (n_s-1) \big[-2+\EulerGamma +\ln(-2 k y)\big] \Big\} \ , 
\end{equation*}
where $\EulerGamma\simeq 0.577$ is the Euler-Mascheroni constant.
Whatever $y$-evolution the remaining terms in Eq. \eqref{eq:2pf2} have
in \textit{this} limit, 
they should precisely cancel the $y$-dependence of the Hankel function.
This is a requirement imposed by conservation of $\zeta$ on super-horizon
scales in single-field models.\footnote{The time-independence of the 
							power spectrum in single-field inflation was
							carefully explained in Ref. \cite{Burrage:2011hd}
							to which we refer the reader for more details.}
We can therefore write the power spectrum as
\begin{equation}
P(k)= \dfrac{H_\star^2\, (1-\varepsilon_\star -s_\star)^2}{4\, z_\star \, (k\, c_{s\star})^3}
\Big\{1+2E_\star + (n_s-1) \ln(k/k_\star) \Big\}\ \ ,
\label{eq:powerspectrum}
\end{equation}
where we have defined 
\begin{equation}
E_\star=-\beta_\star +
\dfrac{n_s-1}{2} \,\big[-2+\EulerGamma +\ln 2 \big]\ \ .
\end{equation}
This is the power spectrum for a quasi scale-invariant Horndeski 
theory. To recover the perfectly scale-invariant formula
obtained by Khoury \& Piazza \cite{Khoury:2008wj},
we simply set $E_\star$ and $n_s-1$ to vanish, since 
these enter the power spectrum as next-order corrections only.
Whilst being time-independent, 
Eq. \eqref{eq:powerspectrum} correctly reproduces the 
expected logarithmic scale-dependence obtained in the
limit when the slow-roll regime is applicable, 
in which $\varepsilon$ and $s$ are slow-roll parameters.
Indeed, we can check that defining the dimensionless version of the power 
spectrum using the usual rule $\mathcal{P}=k^3 P(k)/2\pi^2$, 
Eq. \eqref{eq:powerspectrum}
obeys
\begin{equation*}
n_s-1 =\dfrac{\d \ln \mathcal{P}}{\d \ln k}\ \ .
\end{equation*}
Again, the explicit expression for $n_s-1$ will not be required 
in any stage of the calculation, but we refer the reader to 
appendix \ref{app:ns} where its formula is derived. 
It suffices to treat $n_s-1$ as an arbitrary next-order
quantity.

\subsection{Obtaining next-order corrections to the wavefunctions, $\zeta_k$}
\label{sec:propagator}
Our ultimate interest lies in obtaining the bispectrum of perturbations 
starting from the
action \eqref{eq:zeta-action3}. 
We will apply the same procedure defined in the previous sections 
and rely on the scale-invariant approximation. 
Since we will keep terms to next-order in scale-invariance, we also require corrections 
to the elementary wavefunctions up to the same order, for consistency. 
Next-order corrections to the leading order bispectrum were 
first enumerated by Chen \etal\, for $P(X,\phi)$ models  \cite{Chen:2006nt}.
It is easier to think about these corrections by locating 
them in a Feynman diagram corresponding to the 
three-point function we want to evaluate.
The sources of next-order corrections are therefore quite well-defined and we
explore them in what follows. 

\vspace*{0,3cm}

\para{External legs}
The corrections to the external lines of the Feynman diagram
correspond to evaluating the propagator in the asymptotic regime, 
when $|ky|\rightarrow 0$.
From Eq. \eqref{eq:powerspectrum} we can read off the corrections 
to the external 
legs
\begin{equation}
 \zeta_k(y)^{\textrm{(background+external)}}= \dfrac{\im}{2}
 \dfrac{H_\star (1-\varepsilon_\star-s_\star)}{\sqrt{z_\star} (k\, c_{s\star})^{3/2}}
 \Big\{1+E_\star +\dfrac{n_s-1}{2} \ln (k/k_\star) \Big\} \ \ .
 \label{eq:external legs}
\end{equation}
As argued in \S\ref{sec:power spectrum} the time-independence of this value
is guaranteed because the primordial perturbation 
$\zeta$ is conserved on super-horizon scales in single-field models.

From the requirement of time-independence of the power spectrum, 
we also note that 
whatever function we assign to $z(y)$, it must generically behave, at
leading order in scale-invariance, as
\begin{equation}
z(y) \sim (-k_\star y)^{\frac{2\varepsilon_\star+3s_\star}{1-\varepsilon_\star - s_\star}}\ \ .
\label{eq:z}
\end{equation}
This imposes a mild requirement on an otherwise arbitrary, but smooth 
function $z(y)$. We check that for $P(X,\phi)$ models, where the formula for $z$ is 
well-defined, $z=\varepsilon/c_s^2$, this behaviour is 
consistent with 
Eqs. \eqref{eq:sr1} and \eqref{eq:cs}, together with the condition
$s_\star=-2\varepsilon_\star$ (which holds at leading order in 
the scale-invariant 
approximation).
This formula could not have been anticipated without performing 
the calculation of the power spectrum.
If in a given model $z$ has a different evolution from that of Eq. \eqref{eq:z}, 
it leads to a background where $\zeta$ is evolving on super-horizon
scales, which is incompatible with the single-field inflation scenario.

\vspace*{0,3cm}

\para{Internal legs}
This correction results from the spectrum of scalar perturbations
being slightly tilted. This manifests directly in Eq. \eqref{eq:zeta_k}
for the elementary wavefunction. Using Eq. \eqref{eq:a}, we rewrite $\zeta_k$ for clarity
as
\begin{equation}
\zeta_k (y)=\sqrt{\dfrac{\pi}{8}} \dfrac{H(y) (-y)^{3/2} (1-\varepsilon -s)  }{\sqrt{z(y)}\, c_s^{3/2}\,(y)}
\Big\{ 1-\dfrac{\varepsilon \eta+ts}{(1-\varepsilon -s)^2}  \Big\} \,H_\nu^{(1)}(-ky) \ \ .
\label{eq:zetak2}
\end{equation}
Since the order of the Hankel function, $\nu$, 
differs from $3/2$ by next-order terms, 
these generate corrections to $\zeta_k$. To evaluate these we
start by Taylor expanding the Hankel function around order $\nu_0=3/2$.
Following Refs. \cite{Chen:2006nt, Burrage:2010cu}, we find that the background
evolution of the wavefunctions is given by 
\begin{equation}
 \zeta_k (y)^{\textrm{(background)}}=
 \dfrac{\im H_\star (1-\varepsilon_\star -  s_\star)}{2\sqrt{z_\star}\, (k c_{s\star})^{3/2}}
\ (1-\im ky)\, e^{\im ky}\ \ ,
 \label{eq:backzeta}
\end{equation}
which agrees with the leading order behaviour of Eq. \eqref{eq:external legs}
in the limit $|ky|\rightarrow 0$.
The next-order corrections arising from 
terms proportional to $\delta \nu=\nu-3/2=-(n_s-1)/2$ and 
the slow-variation of the remaining background quantities in Eq. \eqref{eq:zetak2}, 
organise themselves in what we label as internal leg corrections, as follows:
\begin{equation}
		\begin{split}
\delta \zeta_k(y)^{\textrm{(internal)}}=
\dfrac{\im H_\star (1-\varepsilon_\star -s_\star)}{2\sqrt{z_\star}\, (k c_{s\star})^{3/2}}
\bigg\{ &
-e^{\im ky} (1-\im ky) 
\bigg[\beta_\star +\dfrac{n_s-1}{2}\ln(-k_\star y) \bigg]\\
&+\dfrac{n_s-1}{2} 
\bigg[
e^{-\im ky} (1+\im ky) \int_{-\infty}^{y}\dfrac{e^{2\im k \xi}}{\xi} \d \xi -2e^{\im ky} 
-\im \dfrac{\pi}{2} e^{\im ky} (1-\im ky) 
\bigg]
\bigg\} \ .
		\end{split}
		\label{eq:zetakinternal}
\end{equation}
The integral representation on the second line 
of the previous equation corresponds to the exponential 
integral function, $\mathrm{Ei} (-ky)$, defined for real, and non-zero argument:
\[\mathrm{Ei} (x) = \int_{-\infty}^{x} \dfrac{e^{t}}{t}\, \d t \ \ .  \]
The time-dependence of the internal leg corrections
is fairly
intricate and
we verify that there are no divergences in Eq. \eqref{eq:zetakinternal}
when we take $|ky|\rightarrow 0$.
This represents a minimal check on this result. 
Expressions for the time derivatives of the wavefunctions can 
be found in appendix \ref{app:propagator}.

\vspace*{0,3cm}

\para{Corrections arising from evolving interaction vertices} 
The interaction coefficients $\Lambda_i$ can generically 
evolve very fast, and one expects their time evolution to be of the form
\begin{equation}
\Lambda_i \simeq \Lambda_{i\star} (-k y)^{n} \bigg\{1+ \tilde{a} \ln (-k_\star y)
+\tilde{b} \big(\ln (-k_\star y) \big)^2 \bigg\}\ \ ,
\label{eq:fastLambda}
\end{equation}
with the power $n$ depending on the exact expression of $\Lambda_{i\star}$,
and $\tilde{a}$ and $\tilde{b}$ being next-order terms.
This power-law dependence needs to be taken into account 
for consistency with the scale-invariant approximation---we 
will study an explicit example in \S \ref{subsec:features}. 
There we explain in detail that this power-law behaviour can potentially bring 
problems to the convergence of the integral required
to compute the bispectrum
(as we briefly review in \S\ref{sec:bispectrum}). In particular, 
there might be values of $\varepsilon$ for which one is unable to 
perform the calculation, because they would lead to a 
time-dependent three-point function.
Mainly to simplify our results, 
we assume the interaction vertices have a very smooth and slow 
evolution in $y$-time. To this end,
we introduce a supplementary slow-variation 
parameter
\begin{equation}
h_i \equiv \dfrac{1}{\Lambda_i H}\dfrac{\d \Lambda_i}{\d t}\ \ ,
\end{equation}
which satisfies $|h_i|\ll 1$. We expect this approximation to be valid
for all Horndeski models whenever the interaction vertices are 
slowly evolving in time. 
Nevertheless, as described in \S \ref{subsec:features} the methods of this paper can 
still be used to compute $n$-point correlation functions in 
models where this assumption fails.

We conclude that, if $\Lambda_i$ is slowly varying, then
each interaction vertex evolves as 
\begin{equation}
\Lambda_i \simeq \Lambda_{i\star} 
\bigg\{1-\dfrac{h_{i\star}}{1-\varepsilon_\star -s_\star} \ln (-k_\star y) \bigg\}\ \ .
\label{eq:vertex corrections}
\end{equation}
This means that in the remainder of the paper 
we will set, for simplicity, $n=0$ in Eq. \eqref{eq:fastLambda}, but 
we will show how to deal with $n \neq 0$ in one simple example
in \S\ref{subsec:features}.
This assumption concludes the presentation of the 
slow-variation catalogue, which is therefore composed by the set 
$\{n_s-1 , \delta ,\eta, t, h_i \}$.

\para{Further simplifications}
We note that an additional simplification to the background 
dynamics follows from inspection of Eqs. \eqref{eq:external legs}
and \eqref{eq:zetakinternal}. The variable $w$, 
that parametrizes the variation of $z$, does not appear explicitly in our formulae.
Indeed $w$ is only present 
implicitly since it is encapsulated in the spectral index 
$n_s-1$ [cf. Eq. \eqref{eq:qsi_relation} of appendix \ref{app:ns}].
 Therefore the calculation is \textit{independent} of our choice of 
$w$.
As a consequence, we can 
effectively reduce the relation \eqref{eq:kpsimple} between $\varepsilon$, $s$
and $w$, to one which only involves the first two variables. 
To this end, we take the relation between $\varepsilon$ and $s$ to 
be 
\begin{equation}
 s=-2\varepsilon+\delta \ \ ,
 \label{eq:kpsimpler}
\end{equation}
where $\delta$ is a next-order parameter in the scale-invariant approximation. 
We emphasize that 
this procedure in no way lacks generality: the calculation is still 
non-perturbative in $\varepsilon$ and $s$.
The formula above generalizes the regime studied in Ref. \cite{Khoury:2008wj} by considering
a nearly scale-invariant spectrum.
This also implies that $t=\eta$ in the scale-invariant approximation, 
which is in agreement with the observations in \S\ref{sec:dynamics}.
This procedure dramatically simplifies our formulae 
and reduces the slow-variation catalogue to a minimum
of parameters, which we choose to be $\{n_s-1, \delta, \eta, h_i \}$.
At the same time, our results will be displayed in terms of 
only one non-perturbative parameter, which we choose to be $\varepsilon$.\footnote{
							Our calculation is non-perturbative in \textit{both} $\varepsilon$
							and $s$. By realizing that we can 
							write $s=-2\varepsilon +\delta$, 
							we can replace the non-perturbative
							parameter $s$ by the perturbative parameter $\delta$. 
							In the remainder of the text, the dependence in 
							$s$ is absorbed by $\delta$.
				}
We note that this will imply some conceptual restructuring of our formulae.
In particular, using Eq. \eqref{eq:kpsimpler}, 
Eq. \eqref{eq:backzeta} now contains leading 
\textit{and} next-order contributions in $\delta_\star$, which can 
now be interpreted as an extra next-order contribution to the 
external legs 
in Eq. \eqref{eq:external legs}.

Eqs. \eqref{eq:external legs}, \eqref{eq:zetakinternal}, and 
\eqref{eq:vertex corrections} 
are assembled to produce the overall 
corrections which are required to write
the bispectrum at next-order in the scale-invariant approximation.
 They naturally combine 
to produce two distinct contributions to the bispectrum:
the background contributions and next-order corrections 
arising from the vertex and external legs; and 
the next-order contributions arising from the internal legs
of the Feynman diagram. We will use this way of partitioning the bispectrum when presenting
our results in \S\ref{subsec:bispectrum},
and we will label the first contributions as \textit{type a bispectrum}
and the second as \textit{type b bispectrum}.

\section{Non-gaussianity in the bispectrum}
\label{sec:bispectrum}
Non-gaussian signatures encoded in the CMBR work
as a powerful discriminant between inflationary models. 
Quantum field theory correlation functions were initially 
studied by Schwinger \cite{Schwinger:1960qe} and 
Keldysh \cite{Keldysh:1964ud} in the 60s, 
and later applied to Cosmology by 
Jordan \cite{Jordan:1986ug} and 
Calzetta \& Hu \cite{Calzetta:1986ey}.
But it was only
with Maldacena's publication in 2002 \cite{Maldacena:2002vr}
that the applications of the \textit{in-in} formalism
to non-gaussianity were made more clear, 
together with a pair of papers 
by Weinberg \cite{Weinberg:2005vy, Weinberg:2006ac}.
This formalism is the appropriate construction 
to compute expectation values, 
and we briefly summarize its basic ideas here
 (there are a number of papers which review the 
 \textit{in-in} formalism---see, for example, Refs. 
\cite{Seery:2006vu, Koyama:2010xj}). 

In this paper, 
we are interested in computing three-point correlation functions,
$B$, 
at tree-level in the interactions of the comoving curvature 
perturbation, $\zeta$, defined as
\[ \langle \zeta (\vec{k}_1) \zeta (\vec{k}_2) \zeta (\vec{k}_3)  \rangle= 
(2\pi)^3 \delta (\vec{k}_1+\vec{k}_2+\vec{k}_3)   \ 
B(k_1, k_2, k_3) \ \ . \]

In terms of the Argand diagram in complex time $y$, 
these correlations are obtained by performing a path integral from the true
vacuum of the theory, at $y\rightarrow -\infty$, to the time 
of interest when we compute the expectation value. To this we
add the path integral performed backwards 
to $y \rightarrow -\infty$. Schematically, this can be translated into
\begin{equation}
\langle \zeta (\vec{k}_1) \zeta(\vec{k}_2) \zeta(\vec{k}_3)\rangle
= \int \big[\d \zeta_+ \d \zeta_- \big]\ 
\zeta_+ (\vec{k}_1) \zeta_+(\vec{k}_2) \zeta_+(\vec{k}_3)\ 
e^{\im S[\zeta_+]-\im S[\zeta_-]}\ \ ,
\label{eq:correlation}
\end{equation}
where the forward path integral is labelled by the fields $\zeta_+$,
whereas the backwards path integral is labelled by the fields 
$\zeta_-$.\footnote{In practice, to project onto the true vacuum 
				of the interacting theory, one needs to translate the integration contour
				to slightly above and below the negative real $y$-axis, respectively.
				This prescription is in many ways similar to the $i\varepsilon$ trick 
				recurrent in quantum field theory and guarantees convergence of the 
				integral.}
These fields are constrained to agree at any one time \textit{later} than 
that of the observation, and so the contour of integration 
will necessarily turn and cross the real $y$-axis.
From this construction we see that we need the 
cubic action, $S^{(3)}$, for the 
perturbation $\zeta$, written in Eq. \eqref{eq:zeta-action3}, 
to obtain the lowest order non-vanishing Wick contractions between the fields. 
This is because at leading order in the interaction, the first 
non-vanishing contribution arises from contracting the three fields inside the
correlator (essentially three copies of $\zeta$) 
with the three fields in the cubic action.
Since the action is composed by \textit{five} Horndeski operators, 
the total bispectrum will be given by the sum 
of the corresponding \textit{five} (individual) bispectra.
The form of Eq. \eqref{eq:correlation} also 
ensures that the contribution from 
the backwards path integral is precisely the complex conjugate of that 
corresponding to the forward integral.\footnote{This is only true at tree-level for 
					diagrams which involve one interaction vertex only.}
 Therefore, it suffices to compute
one of these path integrals.

There are several ways of presenting the bispectra.
Some authors prefer to present their answers in terms of a 
\textit{rescaled bispectrum},
traditionally defined as \cite{Komatsu:2001rj, Lyth:2005fi}
\begin{equation}
\fNL = \dfrac{5}{6} \dfrac{B(k_1, k_2, k_3)}{P(k_1) P(k_2) + \textrm{cyclic permutations}}\ \ .
\label{eq:fNL}
\end{equation}
Cosmological constraints for $\fNL$ are quoted in Refs. 
\cite{Senatore:2009gt, Curto:2011zt}. For a perfectly symmetrized result for $\fNL$,
we need to add two symmetrized copies of the 
power spectrum---this is what we mean by adding ``cyclic permutations.''
To place constraints in model building, 
it is common to write $\fNL$ evaluated in  
some specific momenta configuration, such as the
equilateral limit for example. We then compare this estimate with the amplitude of the 
bispectrum in the equilateral template used in CMB analysis. 
However, simply because some bispectrum shape peaks in 
the equilateral limit, does not mean it perfectly matches
the bispectrum equilateral template.
Hence, in order to place constraints on the parameters
of some theory, we ought to further compute
the correlation between 
the bispectrum we are studying and its closest template---the error bar associated with the 
constraints should be comparable to the resolution of the data.

Other authors have observed that the entire bispectrum shape
can be used as the individual signature of each inflationary model 
\cite{Fergusson:2008ra, Fergusson:2009nv}.
Decoding the entire bispectrum shape into primitive or fundamental harmonic
shapes
by performing a partial-wave expansion
acts in the same way as 
the decomposition of the angular power spectrum in $C_\ell$'s, 
once we identify the number of the harmonic with the multipole
moment, $\ell$.
Whatever final bispectrum shape is generated in a given model, 
it will 
be a linear superposition of 
five individual bispectra, 
with weights given by the respective interaction 
coefficients, $\Lambda_i$.
This has a major significance for the overall bispectrum shape: 
it can be viewed as a unique fingerprint of the Lagrangian structure of the theory.
In this perspective, there is no real advantage in specifying 
the results for $\fNL$ when one can use the entire bispectrum shape as the most
sensitive and discriminating degree of freedom.

Which of these
		methods will prove to be more 
		efficient in data analysis
		is not absolutely clear at the moment, 
		although the modal 
		decomposition seems to be 
		quite promising (see, for example, 
		Ref. \cite{Regan:2011zq} for comments 
		on the efficiency of this algorithm). 
		In particular, with this approach
		it is possible to derive consistency relations 
		between the amplitudes of 
		each harmonic, which make it easy to 											
		rule out classes of inflation models 
		whenever these relations are not supported by the data \cite{Ribeiro:2011ax, Battefeld:2011ut}.
In this paper we shall adopt this last philosophy and present the general 
formulae for the bispectrum. The study of shapes is, 
however, beyond the scope of this paper, 
and we leave it to forthcoming work.
We include estimates of $\fNL$ in \S\ref{subsec:features} in specific applications only.

\subsection{Bispectrum of Horndeski theories}
\label{subsec:bispectrum}

As argued before, to compute the bispectrum it will 
be more convenient to write the action for perturbations in 
$y$-coordinates. From Eq. \eqref{eq:zeta-action3} this is given by
	\begin{equation}
		S^{(3)} = 
		\int \d^3 x \, \d y\, a^2 \; \bigg\{ 
			\frac{c_s^2}{a} \Lambda_1 \zeta'^3 + c_s\Lambda_2 \zeta\,\zeta'^2
			+ \frac{1}{c_s} \Lambda_3 \zeta (\partial \zeta)^2
			+  c_s\Lambda_4 \zeta' \partial_i \zeta
				\partial^i (\partial^{-2}\zeta')
			+  c_s \Lambda_5 \partial^2 \zeta
				(\partial_i \partial^{-2}\zeta')^2	
		\bigg\} \ ,
		\label{eq:zeta-action3y}
	\end{equation}
	where we recall that primed operators are differentiated with respect to $y$.
Our calculation is organised as follows.

\vspace*{0,3cm}

\para{Bispectrum type $a$}
The leading contributions to the bispectrum 
and the corrections arising from the slow-variation 
of the interaction vertices and external legs can be 
written in the general form
\begin{equation}
B^a =\dfrac{\Lambda_{i\star}\, H_\star^4 \, (1+\varepsilon_\star)^4 }{2^6 c_{s\star}^6 z_\star^3 \prod_i k_i^3}\, N^a (k_1)\, \bigg\{
\Re\Big(P^a(k_1) \tilde{J}_\gamma \Big)+
\Re\Big(Q^a(k_1) \tilde{J}_{\gamma+1} \Big) +T^a(k_1)
  \bigg\} \ + \ \textrm{cyclic permutations}\ \ ,
  \label{eq:Ba}
\end{equation}
where the addition of cyclic permutations entails the symmetric exchange
of momenta $k_1 \rightarrow k_2 \rightarrow k_3$. The functions 
$N^a(k_1)$, $P^a(k_1)$, $Q^a(k_1)$ and $T^a(k_1)$ 
are listed on table \ref{table:bafunctions}.
The functions $\tilde{J}_\gamma$ are studied in 
appendix \ref{app:other_integrals} and defined by

\begin{equation}
\tilde{J}_\gamma \equiv  (\im k_t)^{1+\gamma} \int_{-\infty}^{0} \d y e^{\im k_t y} \ (-y)^\gamma 
\bigg\{
A_\star + B_\star \ln (-k_\star y)+C_\star \big(\ln (-k_\star y) \big)^2 
\bigg\}\ \ .
\label{eq:Jtildeintegral}
\end{equation}
Table \ref{table:JcoefficientsBa} contains the assignments $A_\star$, $B_\star$
and $C_\star$ for the functions
$\tilde{J}_\gamma$,
and we note that the coefficients 
for the operators $\zeta \zeta'^2$, 
$\zeta' \partial_j \zeta \partial_j \partial^{-2} \zeta' $, 
and $\partial^2 \zeta (\partial_j \partial^{-2} \zeta')^2$
are the same. Having done the calculation independently for 
each operator, this is a minimal check of the correctness of our results.
This is because these operators have the same time derivative structure, and differ
only in the arrangement of spatial derivatives, and therefore, in the momentum 
dependence.

\vspace*{0,3cm}

\para{Bispectrum type $b$}
Likewise, the contributions from the propagator in the 
internal lines of the Feynman diagram can be consolidated
for all operators in the form
\begin{equation}
		\begin{split}
B^b =\dfrac{\Lambda_{i\star}\, H_\star^4 \, (1+\varepsilon_\star)^4 }{2^6 c_{s\star}^6 z_\star^3 \prod_i k_i^3}\, N^b(k_1)\, \bigg\{ & (n_s-1)
\sum_i\Big[\,\Re\Big(P^b_i(k_1) \Big)\tilde{I}_\gamma (k_i) +
\Re\Big(Q^b_i(k_1)\Big) \tilde{I}_{\gamma+1}(k_i)  \Big]\\
&+\Re \Big(R^b(k_1) \,\tilde{J}_\gamma\Big)
+\Re \Big(S^b(k_1)\, \tilde{J}_{\gamma+1}\Big)
+(n_s-1)\, T^b(k_1)
  \bigg\} \  \\ & + \textrm{cyclic permutations}\ \ ,
  \label{eq:Bb}
  \end{split}
\end{equation}
where the addition of cyclic permutations entails the symmetric exchange
of momenta $k_1 \rightarrow k_2 \rightarrow k_3$. The functions 
$N^b(k_1)$, $P^b_i(k_1)$, $Q^b_i(k_1)$, $R^b(k_1)$, $S^b(k_1)$ and 
$T^b(k_1)$ are listed on tables \ref{table:bbfunctions1}
and \ref{table:bbfunctions2}.
Table \ref{table:IJcoefficientsBb} contains the assignments 
$A_\star$, $B_\star$
and $C_\star$ for the functions
$\tilde{J}_\gamma$.
The functions $\tilde{I}_\gamma (k_i)$ are defined by
\begin{equation}
\tilde{I}_\gamma (k_3) \equiv 
(\im \vartheta_3 k_t)^{\gamma+1}
\int_{-\infty}^0 {\d y}\, \bigg\{  {(-y)^\gamma}  
e^{\im (k_1+k_2-k_3) y}\, \int_{-\infty}^{y}{\dfrac{\d \xi}{\xi}\, e^{2\im k_3 \xi}} 
\label{eq:Itilde}
\bigg\}\ \ ,
\end{equation}
where $\vartheta_3= (k_t-2k_3)/k_t$ (we refer the reader to appendix \ref{app:integral_function} 
for details on the evaluation of these functions).

\vspace*{1cm}
Both types of corrections are computed using the same
\textit{in-in} formalism rules. The third Horndeski operator, $\zeta (\partial \zeta)^2$, 
however, presents an additional degree of complexity (which also had 
to be dealt with in Ref. \cite{Burrage:2011hd}). We briefly revisit it here. 
We start by noting that this operator is undifferentiated (with respect 
to time), and therefore the integrand is, at least, power-law divergent. 
This is because there are insufficient powers of $y$ in numerator, 
to counteract the powers in denominator 
owing to the presence of $a(y)$ in the integrand. As a result 
the integrand behaves, at leading order, 
like $a^2/c_s \sim 1/y^2$.\footnote{The calculation of the bispectrum
						of this operator using the scale-invariant approximation
						 is very similar to the one 
						presented in Ref. \cite{Burrage:2011hd}, which 
						was nevertheless restricted to the slow-roll 
						approximation. This happens because, 
						contrary to the remaining operators, the constant 
						$\gamma$ in the integrals in Eqs. \eqref{eq:integfunctionint}
						and \eqref{eq:Jintegral}
						is an integer, which allows us to
						write compact results in tables 
						\ref{table:bafunctions} and \ref{table:bbfunctions1}
						for this operator.}
Moreover, the next-order corrections in the scale-invariant approximation add 
logarithmic terms to the integrand function.
This 
has serious repercussions in the final result because it can
potentially lead to 
power-law and logarithmic divergences in $y$ when one takes the 
limit $|y|\rightarrow 0$---these are \textit{dangerous interactions}, as 
named by Weinberg \cite{Weinberg:2005vy}. In order to ensure that 
the real part of the correlation function converges in the asymptotic limit, 
one needs to carefully isolate the primitively divergent contributions:
either power-laws like $|y|^{-\alpha}$, with $\alpha>0$, or logarithmic
terms, $\ln (-y)$. 
This is done by integrating all divergent integrals by parts
an appropriate number of times. 
Only two situations occur as a result:
\begin{enumerate}[i.]
\item the isolated divergent terms contribute with a purely 
imaginary part to the correlation function. In this case, this divergence
becomes irrelevant since the total correlator is given 
by the sum of two path integrals which are complex conjugates
(yielding a purely real final result).
\item the isolated divergent contributions
are real and contribute to the final answer. However,
when we sum \textit{type a} and \textit{type b bispectra}, these divergent 
contributions 
cancel out
amongst themselves. The final result contains only
finite contributions.
\end{enumerate}
It is crucial to take into account these two possibilities 
to obtain a correct result:
first, to ensure that the correlation functions do not diverge 
in the asymptotic limit $|ky|\rightarrow 0$; and second, to guarantee that the calculation 
contains all the relevant convergent contributions to the overall 
bispectrum. This is particularly important when checking whether 
our results are consistent with Maldacena's factorization theorem
\cite{Maldacena:2002vr, Creminelli:2004yq}---this asserts that, in the limit when one of the 
momenta is small, the bispectrum should factorize into two copies of 
the power spectrum multiplied by $n_s-1$; in other words, 
in this limit, two and three-point correlators talk to each other.

\subsection{Features of the bispectrum of Horndeski theories}
\label{subsec:features}
These results extend those obtained assuming 
a slow-roll inflationary phase in the early Universe.
Some comments about our formulae share the same 
ideas as in previous works in the literature, and we summarize them 
here for completeness.
\vspace*{0,3cm}

\para{Enhancement of non-gaussianities} It is apparent from 
our formulae that the enhancement of the bispectrum can be 
very different from the one found in models  
assuming the slow-roll approximation. 
This strictly depends on the expressions of the coefficients $\Lambda_i$. 
As observed in Ref. \cite{Burrage:2010cu}, if the interaction vertices
in the action \eqref{eq:zeta-action3y} 
do not contain enough powers in the speed of sound, 
then the overall dependence in $c_s$ can be stronger than 
that commonly associated with DBI models $\fNL \sim c_s^{-2}$.
Depending on $z$, more exotic models could  
reproduce such behaviour. To gauge whether this scenario would 
be permissible could involve applying the partial wave decomposition 
method to the Horndeski overall bispectrum shape mentioned in \S\ref{sec:bispectrum}. 
Five measurements 
would be required to fix each of the $\Lambda_i$ interactions, 
and a number of additional measurements to break the degeneracy in 
other parameters, such as $\varepsilon$ and $c_s$.
Ultimately this would allow us to place constraints on all the 
parameters of the theory. Only then would we be able to conclude 
whether there is enhancement of non-gaussianities in these models.

\vspace*{0,3cm}

\para{Presence of logarithms} Our formulae contain logarithms
of momenta encoded, for example, in the master integral $J_\gamma$ 
(defined in Eq. \eqref{eq:Jintegral} of appendix \ref{eq:integfunctionint2}). 
There are two varieties of logarithms
as noted in Ref. \cite{Burrage:2011hd}: those which depend on the reference scale, 
$\ln (k_\star/k_t)$, and those which 
depend on the perimeter momentum scale, $\ln (k_i/k_t)$.
The first type of logarithms are clearly of the same nature as the 
ones identified in the power spectrum \eqref{eq:powerspectrum}: 
they encode the scale-dependence of the bispectrum.
One can choose the reference scale, $k_\star$, 
so as to minimize these; alternatively, one can use this degree 
of freedom to measure primordial non-gaussianities on different scales.
The last type of logarithms are shape dependent. To better understand this,
one can write one of the three momenta, $k_i$, in terms of the perimeter, 
$k_t$, and two additional coordinates, describing the angular
dependence. This implies that $\ln(k_i/k_t)$ is effectively only
a function of the angular coordinates, and it is therefore responsible 
for the shape-dependence of the bispectrum.

\vspace*{0,3cm}

\para{Away from slow-roll}
If the inflationary background is almost de Sitter,
so that $\varepsilon \ll 1$, 
it is easy to see that the power-law  behaviour
$k_\star^{\alpha \frac{\varepsilon_\star}{1+\varepsilon_\star}} $, 
where $\alpha$ is some integer, can be written as a next-order
logarithmic
correction by performing a Taylor series expansion. 
Indeed by taking the limit of 
very small $\varepsilon$, our results 
agree with those of Ref. \cite{Burrage:2011hd}.
Away from the slow-roll regime, 
when the slow-roll approximation is no longer applicable, 
the dependence on the 
reference scale, $k_\star$, arises from power-laws 
(whose Taylor expansion we cannot truncate)
in addition to logarithmic contributions. 

For comparison, let us write the formula of the \textit{bispectrum type a} 
for one operator, say $\zeta'^3$. We find that:
\begin{equation}
\begin{split}
B^{(a)}_{ \zeta'^3} = 
\dfrac{12 H_\star^5\, (1+\varepsilon_\star) ^5 \,\Lambda_{1\star} }{2^6 \, z_\star^3\, c_{s\star}^6  \,k_t^3\, \prod_i k_i}\bigg(\dfrac{k_\star}{k_t}\bigg)^{\frac{5\varepsilon_\star}{1+\varepsilon_\star}}\, \Gamma(\xi ) &
\bigg\{ 
\cos(\varsigma) 
 \bigg[
A_\star +  \ln\frac{k_\star}{k_t}\, 
\bigg(B_\star + C_\star\ln \frac{k_\star}{k_t} \bigg)\\
		&  \ \  \ \ \ \ \ \ \ \ \ \ \  \ \,  
		+\psi^{(0)} (\xi) \bigg(
		B_\star +2C_\star \ln\frac{k_\star}{k_t}
		+C_\star \psi^{(0)} (\xi)
		\bigg) + C_\star \psi^{(1)} (\xi) \bigg]\\
&-
\frac{\pi}{2} \, \sin(\varsigma) 
 \bigg[
B_\star + C_\star\ln \frac{k_\star}{k_t} 
+ \ln\frac{k_\star}{k_t}\, 
+2C_\star\psi^{(0)} (\xi) \bigg] \ \bigg\}\ \ \ ,
\end{split}
\label{eq:op1}
\end{equation}
with $\xi\equiv 3+\frac{5\varepsilon_\star}{1+\varepsilon_\star}$ and
$\varsigma \equiv \frac{5\pi \varepsilon_\star}{2(1+\varepsilon_\star)}$.
This expression exhibits an unusual power-law dependence on the reference
scale, $k_\star$, which is absent from previous studies, 
where only the logarithmic contributions $\ln(k_\star/k_t)$
were known \cite{Burrage:2011hd}. 
By keeping the reference scale $k_\star$ arbitrary in our 
calculation, the scale-dependence of the bispectrum can be calculated. 
This will include the contribution of the power-law scaling as
$k_t^{-3-\frac{5\varepsilon_\star}{1+\varepsilon_\star}}$, 
which could receive large corrections from $\varepsilon_\star$.
In Ref. \cite{Khoury:2008wj} Khoury \& Piazza chose $k_\star=k_t$, 
which masks the power-law effect.

Considering a slow-roll expansion, by which 
we take $\varepsilon_\star$ to be a slow-roll 
parameter (treating it on equal footing as $\delta$ and $n_s-1$), 
Eq. \eqref{eq:op1} simplifies to
\begin{equation}
\begin{split}
B^{(a)}_{ \zeta'^3} \simeq \dfrac{3\Lambda_{1\star}H^5_\star}{8 z_{\star}^3 c_{s\star}^6 \prod_i k_i k_t^3} 
\bigg\{ & 1+ \frac{h_{1\star}}{2} (-3+2\EulerGamma) +\frac{3 (n_s-1)}{2} (-2+\EulerGamma)
+\frac{\delta_\star}{2} (-13+6\EulerGamma) +\frac{\varepsilon_\star}{2} (25-10\EulerGamma)\\
&+\frac{3(n_s-1)}{2} \ln \bigg(\frac{2k_1k_2k_3}{k_\star^3} \bigg)
- (h_{1\star}+3\delta_\star-5\varepsilon_\star) \ln \frac{k_\star}{k_t}
 \bigg\}\ \ ,
 \label{eq:srop1}
 \end{split}
\end{equation}
which resembles the bispectrum obtained in 
Ref. \cite{Burrage:2011hd}.\footnote{To be more precise, 
			we can indeed show that our results 
			for the overall bispectrum agree (summing\textit{ type a} and \textit{type b}), 
			including the explicit dependence on the reference scale, $k_\star$.} 
We note that in Eq. \eqref{eq:srop1} the dependence on $k_\star$ appears 
through logarithms, and the power-law behaviour has 
explicitly disappeared from the result. 
As described above, this is because by taking
$\varepsilon$ to be a small parameter, the power-law is 
well described, at next-order in the slow-roll approximation, by a logarithmic contribution.
We also observe that the reference scale, $k_\star$, appears
in the form of two logarithmic functions: $\ln (k_i/k_\star)$
and $\ln (k_\star/k_t)$. These species of logarithms 
were thoroughly studied in Ref. \cite{Burrage:2011hd}
to which we refer the reader for details.
It is in this sense that our work generalizes the calculations which
have been performed assuming the slow-roll approximation.

The behaviour described above shows that 
whereas the power spectrum has a universal weak, logarithmic scale-dependence 
[cf. Eq. \eqref{eq:powerspectrum}], 
the bispectrum of these theories beyond the slow-roll regime can have a
 much stronger scale-dependence. 
In principle, this could be used to distinguish, 
on observational grounds, between the 
slow-roll and scale-invariance regimes: from 
CMB ($k \lesssim 0.5 h \mathrm{Mpc}^{-1}$) to cluster scales 
($k \gtrsim 0.5 h \mathrm{Mpc}^{-1}$), 
interpolating with scales probed by the galaxy bispectrum.
A number of authors have studied 
constraints on non-gaussianity arising from 
galaxy surveys \cite{Matarrese:2000iz, Bernardeau:2001qr}, 
including its relation with biasing \cite{Verde:1999ij}. 
The scale-dependence of the bispectrum was initially studied 
by Chen in an infrared model of DBI inflation \cite{Chen:2005fe}, 
and later investigated in $P(X,\phi)$ models 
(a subset of the Horndeski class) by LoVerde \etal\,\cite{LoVerde:2007ri}.
It is not the aim of this paper to present a detailed analysis 
of the scale-dependence of the bispectrum. We rather want to offer an explicit
example of theories which inherit 
a highly intricate scale-dependence from the background dynamics.

\subsection*{Comparison with previous results}
In certain limits, some of our results overlap with others
in the literature. We compare them here.
\vspace*{0,3cm}

\para{Khoury \& Piazza}
First, we recall that Khoury \& Piazza  \cite{Khoury:2008wj} have 
calculated the bispectrum of Lagrangians involving only 
the inflaton and its first derivative in an exactly scale-invariant 
background, in which $s_\star=-2\varepsilon_\star$. 
In this paper, we have extended this study in two ways:
by performing 
the calculation perturbatively away from the exact scale-invariance regime
and applying it to the Horndeski class of models. 
In particular, let
us take a specific example for comparison. Focusing on the operator $\zeta \zeta'^2$
and in the action studied by Khoury \& Piazza we take 
\[ \Lambda_2 (y)= \frac{\varepsilon}{c_s^4} \big(\varepsilon -3 +3c_s^2 \big) \ \ .\]
We note that $\Lambda_2(y)$ is rapidly varying in time and 
should therefore be placed inside the integral to compute the bispectrum
[cf. Eq. \eqref{eq:correlation}], as follows
\begin{equation}
 B_{\zeta \zeta'^2} \sim \int_{-\infty}^{0} {\d y } \Lambda_2(y)\,  c_s(y)\, a^2(y)  y^2 (1-\im ky)\ \ ,
 \label{eq:BKP}
\end{equation}
where the factor $y^2$ comes from the two time derivatives of 
the wavefunctions and the remaining 
$y$-dependence from the undifferentiated wavefunction. 
For comparison purposes, we only retain the contributions 
at leading order in scale-invariance, which means setting all the 
next-order
corrections we have focused on this paper to zero.
Therefore, selecting only the leading order contribution
to our bispectrum in Eq. \eqref{eq:Ba}, 
we find 
\begin{equation}
\begin{split}
B^{(\textrm{leading})}_{\zeta \zeta'^2} = 
\dfrac{H_\star^4\, (1+\varepsilon_\star) ^4 \, k_2^2 k_3^2}{16 \, c_{s\star}^4 \, \varepsilon_\star^2 \, \prod_i k_i^3 \, k_t^2} &
\bigg\{ 
3c_{s\star}^2  (k_1+k_t) +\\
&+(\varepsilon_\star -3)
\bigg(\frac{k_t}{k_\star}\bigg)^{\frac{4\varepsilon_\star}{1+\varepsilon_\star}} 
\Gamma\bigg(1-\frac{4\varepsilon_\star}{1+\varepsilon_\star}\bigg) 
\cos\bigg(\frac{2 \pi \varepsilon_\star}{1+\varepsilon_\star} \bigg) 
\big(k_1+k_t +\varepsilon_\star (k_t-3k_1) \big)
\bigg\}\\
&  +\textrm{cyclic permutations}\ \ \ ,
\end{split}
\label{eq:BKP2}
\end{equation}
where we have retained the scale-dependence through $k_\star$. 
This formula reproduces the results of Ref. \cite{Khoury:2008wj}
provided we choose the reference scale to satisfy $k_\star=k_t$.
Eq. \eqref{eq:BKP2} is the bispectrum of the operator $\zeta \zeta'^2$
for a perfectly scale-invariant spectrum of perturbations.

Whenever comparison was possible 
our results reproduce those of Ref. \cite{Khoury:2008wj}.

\vspace*{0,3cm}

\para{Noller \& Magueijo}
Second, in Ref. \cite{Noller:2011hd} 
Noller \& Magueijo estimated 
the magnitude of the next-order corrections
in the scale-invariant approximation to the bispectrum
of
$P(X,\phi)$ models. In comparison, our calculation 
extends to the larger Horndeski scalar field theories.
Their estimate focused on
an approximation of 
a subset of next-order corrections important when $|ky|\rightarrow 0$. 
However, there are other contributions which 
contribute equally to the bispectrum and are 
sensitive to the dynamics around the horizon crossing, 
when the approximation $|ky|\rightarrow 0$ fails. 
As we have shown, these corrections can be obtained by evaluating 
the change in 
the Hankel function of order 
$\nu=3/2-(n_s-1)/2$,
rather than at the exact scale-invariant choice, $\nu=3/2$.

Moreover, in our calculation we have performed a uniform expansion to 
next-order in the scale-invariant
approximation, using Eq. \eqref{eq:scale_factor}
for the evolution of the scale factor, $a$, and Eq. \eqref{eq:cs} for the 
evolution of the sound speed of perturbations, $c_s$. Ref. \cite{Noller:2011hd} 
assumed a perfectly scale-invariant background, 
on top of which perturbations would develop.
This amounts to setting $\alpha_\star=\beta_\star=0$
in our Eqs. \eqref{eq:scale_factor} and \eqref{eq:cs}, 
but their contribution is as important as the terms 
linear in $n_s-1$---as we have previously argued,
the scalar fluctuations are sensitive
to the background dynamics, and in particular 
to the next-order corrections we have calculated
in the scale-invariant approximation.

For completeness, we will investigate
 one particular limit 
 in $P(X,\phi)$ theories (these include DBI inflation) 
 studied in Ref. \cite{Noller:2011hd}
when the speed of sound of fluctuations is small, $c_s\ll 1$. This regime
is phenomenologically interesting since it is known to be
related with large non-gaussianities \cite{Alishahiha:2004eh}, 
for which the dominant operator is $\zeta'^3$. 
To make this application more explicit we apply the notation 
of Ref. \cite{Burrage:2011hd} (see also Ref. \cite{Chen:2006nt}), 
and take the action
\begin{equation}
S^{(3)} \supseteq \int \d ^3 x \, \d y
a\Lambda_1 c_s^2 \zeta'^3\ \ ,
\label{eq:action_NM}
\end{equation}
with 
\begin{equation*}
\Lambda_1= \frac{\varepsilon}{H c_s^4} \bigg(
1-c_s^2 -2c_s^2 \frac{\lambda}{\Sigma}
\bigg)\ \ ,
\end{equation*}
and where
\begin{equation*}
\dfrac{\lambda}{\Sigma} \equiv \dfrac{1}{6}\bigg(
\dfrac{2f_X+1}{c_s^2}-1 \bigg) \ \ .
\end{equation*}
$f_X$ is assumed to be constant.\footnote{In DBI models, $f_X=1-c_s^2$ 
										and we cannot technically
										require $f_X$ to be constant, given that we 
										are interested precisely in the regime when 
										$c_s$ is rapidly varying.
										However, for the purposes of making this comparison
										more transparent, we assume this is the case, 
										similarly to what was done in Refs. 
										\cite{Khoury:2008wj, Noller:2011hd, Burrage:2011hd}.
										Rigorously, a more precise estimate would have to 
										take the time-dependence of $f_X$ into account.}
We can see that taking the limit of small $c_s$, 
corresponds to considering
$\big|\frac{\lambda}{\Sigma} \big| \gg 1$, 
in which case the interaction vertex is well
approximated by
\begin{equation*}
\Lambda_1 (y) \simeq -\dfrac{2}{3} \dfrac{\varepsilon f_X}{H c_s^4}\ \ .
\end{equation*}
Inspection of this formula reveals that this interaction 
vertex is rapidly varying, by virtue of its dependence in $\varepsilon$, 
$H$ and $c_s$, of which only $\varepsilon$ is slowly varying
in the scale-invariant approximation. 

Our formulae are easily adapted to the case of rapidly varying $\Lambda_i$.
For $\Lambda_1$ we find
\begin{equation}
\begin{split}
\Lambda_1 (y) =\Lambda_{1\star} (-k_\star y)^{-\frac{9\varepsilon_\star}{1+\varepsilon_\star}}
 \bigg\{ &
 1-\frac{1}{1+\varepsilon_\star}\bigg[\eta_\star +9 \varepsilon_\star \beta_\star +\frac{\varepsilon_\star \delta_\star}{1+\varepsilon_\star} +\frac{4\varepsilon_\star \delta_\star (\varepsilon_\star -1)}{1+\varepsilon_\star} \bigg] \ln (-k_\star y) \\
 & +\frac{9}{2} \frac{\alpha_\star \varepsilon_\star}{1+\varepsilon_\star} \big(\ln (-k_\star y) \big)^2
 \bigg\}\ \ ,
 \end{split}
 \label{eq:general_Lambda}
\end{equation}
up to next-order corrections in the scale-invariant approximation, 
as described in Eq. \eqref{eq:fastLambda}.
These corrections can be absorbed into the coefficients 
$A_\star$, $B_\star$ and $C_\star$ quoted in table \ref{table:JcoefficientsBa}. 
This generalizes Eq. \eqref{eq:vertex corrections} for
rapidly varying interaction vertices. Since 
Eq. \eqref{eq:general_Lambda} adds power-law contributions to the 
integral in Eq. \eqref{eq:action_NM}, 
one might worry that not all the previously allowed values of $\varepsilon$
will allow convergence of the final result---we recall that the overall power-law
needs to decay faster than $(-y)^{-1}$
for convergence criteria to be met. 
In the event this does not happen, then one is required to 
perform integration by parts an appropriate number of times
so as to isolate the primitively divergent contributions, in a 
completely analogous way as we have dealt with the $\zeta (\partial \zeta)^2$
operator. A convergent answer should similarly place 
constraints on the allowed values of $\varepsilon$ so that 
the correlators do not evolve in time, since this would signal 
a spurious divergence.

Nevertheless, for this operator it turns out that 
$\gamma=2-4\varepsilon_\star/(1+\varepsilon_\star)$ in the integral
\eqref{eq:Jintegral}. The integral is therefore
 always convergent
since the condition $\Re \big(\varepsilon_\star/(1+\varepsilon_\star) \big)<3/4$
is satisfied for all the range of $0<\varepsilon_\star<1$.
The calculation is carried out in a similar fashion to what we described in 
\S\ref{subsec:bispectrum}, and which we take as a basis.
Using the definition of $\fNL$ in Eq. \eqref{eq:fNL},
we find that at leading order in the scale-invariant approximation
\begin{equation}
\fNL^{\textrm{(leading)}} = \dfrac{5 f_X}{c_{s\star}^2} \dfrac{\prod_i k_i^2}{\sum_i k_i^3}
\bigg(\frac{k_\star}{k_t} \bigg)^{\frac{3-\varepsilon_\star}{1+\varepsilon_\star}}
k_\star^{-3} (1+\varepsilon_\star)\, \Gamma (\Upsilon )\, \cos \bigg(\dfrac{\pi (1-\varepsilon_\star)}{1+\varepsilon_\star} \bigg) \ \ ,
\label{eq:nmleading}
\end{equation}
 where $\Upsilon=\frac{3-\varepsilon_\star}{1+\varepsilon_\star}$ is 
 a non-negative constant.
This result gives the
possibility of $\fNL$ changing sign if $\varepsilon_\star >1/3$, when the argument 
of the trigonometric function changes from the first 
to the second quadrant. 
This could be important since WMAP constraints predict predominantly
positive values for $\fNL^{\textrm{(equilateral)}}$, whereas the original 
DBI model gives $\fNL^{\textrm{(equilateral)}}<0$ under the slow-roll approximation \cite{Alishahiha:2004eh}.

Including next-order corrections, the result becomes substantially more
complicated and in an attempt to 
simplify these expressions as much as possible, we 
evaluate $\fNL$ in the so called equilateral limit
$k_1=k_2=k_3=k$, and when $k_\star=k$.
We organize the corrections to the leading order
non-gaussianity, $\delta \fNL$, in terms 
proportional to the various slow-variation parameters, as follows:
\begin{equation}
\delta \fNL^{\textrm{(equilateral)}} = (n_s-1) \fNL^{n_s-1} +
\delta \, \fNL^{\delta} +
\eta  \fNL^{\eta}\ \ ,
\label{eq:deltafNL1}
\end{equation}
with
	\begin{subequations}
		\begin{align}
			\fNL^{n_s-1}
			& =  \dfrac{5 f_X k^{2+\frac{4}{1+\varepsilon_\star}}}{2 c_{s\star}^2 (1+\varepsilon_\star)^2} \, 
			\bigg\{
			3^{-\frac{3-\varepsilon_\star}{1+\varepsilon_\star}} \cos \bigg(\frac{2\pi}{1+\varepsilon_\star}\bigg) k^{-\frac{2(3+\varepsilon_\star)}{1+\varepsilon_\star}}
			(1+\varepsilon_\star)^3\, \Gamma(\Upsilon)\ \mathcal{H} _{\varpi} \nonumber
			\\
			& \hspace*{2,8cm}- (1+\varepsilon_\star)^3\, 3^{\frac{4\varepsilon_\star}{1+\varepsilon_\star}}\,
k^{-\frac{2(3+\varepsilon_\star)}{1+\varepsilon_\star}}	 \Gamma(\Upsilon)
\cos \bigg(\frac{2\pi}{1+\varepsilon_\star} \bigg)  \bigg(2+2\EulerGamma +\ln \frac{27}{2} \bigg) \nonumber		\\
& \hspace*{2,8cm}
		-(1+\varepsilon_\star)^3 k^{-6+\frac{4\varepsilon_\star}{1+\varepsilon_\star}}
		\sin \bigg(\frac{2\pi \varepsilon_\star}{1+\varepsilon_\star } \bigg)
	\mathcal{J} \bigg\}\ \ ,
			\\ 	
	\fNL^{\delta}
	&=	-\dfrac{5 \times 3^{-\frac{4}{1+\varepsilon_\star}}}{2 c_{s\star}^2 (1+\varepsilon_\star)}
	\, f_X \Gamma (\Upsilon)	
	\bigg\{
	\cos \bigg( \frac{\pi}{2}\Upsilon\bigg) \pi \Big(3+\varepsilon_\star (-5+4\varepsilon_\star)\Big) \nonumber\\
	&\hspace*{4cm} 
	+\sin \bigg(\frac{\pi}{2}\Upsilon \bigg) \bigg[
	2\Big[1+\varepsilon_\star +\ln3 \Big(-3+\varepsilon_\star (5-4\varepsilon_\star)\Big) \Big]
	\nonumber \\
	& \hspace*{6.3cm}
	+2 \Big(3+\varepsilon_\star (-5+4\varepsilon_\star)\Big) \psi^{(0)} (\Upsilon)
	\bigg]
	\bigg\} \ \ ,\ \ \textrm{and} \\
	\fNL^{\eta}
	& = \dfrac{5\times 3^{-\frac{4\varepsilon_\star}{1+\varepsilon_\star}}k^{2+\frac{4}{1+\varepsilon_\star}}}{2\,c_{s\star}^2 (1+\varepsilon_\star)^2} \, f_X  \Gamma(\Upsilon) 
	\bigg\{
	\cos \bigg(\frac{\pi \varpi}{2}\bigg)  k^{-\frac{2(3+\varepsilon_\star)}{1+\varepsilon_\star}}
	 \bigg[-2\varepsilon_\star (1+\varepsilon_\star)
	+\ln 3 \Big(2+2\varepsilon_\star (1+2\ln 3)\Big) 
	\nonumber\\
	& \hspace*{5cm} -\Big(2(1+\varepsilon_\star)+ 8 \varepsilon_\star \ln 3
	 -4\varepsilon_\star \psi^{(0)} (\Upsilon) \Big)\psi^{(0)} (\Upsilon)
	 +4\varepsilon_\star \psi^{(1)} (\Upsilon)\bigg] \nonumber\\
& \hspace*{4.5cm} -8\, \cos \bigg(\frac{\pi \varpi}{2}\bigg)\, k^{-3-\frac{3-\varepsilon_\star}{1+\varepsilon_\star}}	
\varepsilon_\star (1+\varepsilon_\star ) 
	\nonumber  \\
	& \hspace*{4.5cm}
	+\frac{1}{2}\sin  \bigg(\frac{\pi \varpi}{2}\bigg) k^{-\frac{2(3+\varepsilon_\star)}{1+\varepsilon_\star}}\bigg[
	\pi +4\varepsilon_\star \pi -8\varepsilon_\star \pi \psi^{(0)} (\Upsilon)
	\bigg]
	\bigg\}\ \ .
		\end{align}
		\label{eq:deltafNL2}
	\end{subequations}
\noindent where $\varpi\equiv  2 (1-\varepsilon_\star)/(1+\varepsilon_\star)$. 
We note the dependence on the scale through $k$, 
expected whenever the slow-roll approximation 
breaks down.
The appearance of the polygamma functions of order zero, $\psi^{(0)}$, 
in our results is discussed in Eq. \eqref{eq:integfunctionint3}
of appendix \ref{app:integrals}. 
Also, $\mathcal{H}_m$
denotes the $m^{\textrm{th}}$-harmonic number which relates to 
$\psi^{(0)}$ via $\mathcal{H}_{m-1}=\psi^{(0)} (m)+\EulerGamma$,
and $\mathcal{J}$ satisfies
\begin{equation}
\mathcal{J}=
\Gamma(\Upsilon)  \bigg[\EulerGamma +\ln 2+\psi^{(0)}(\Upsilon) \bigg]
\nonumber
-2 \Gamma \bigg(\frac{4}{1+\varepsilon_\star}\bigg) \ \ 
_3 F_2\bigg( \bigg\{1,1,\frac{4}{1+\varepsilon_\star}\bigg\}, \{2,2\}, -2 \bigg)\ \ ,
\end{equation}
where we have used the results derived in appendix \ref{app:integrals}. 
Eq. 
\eqref{eq:deltafNL1}
contains all next-order contributions in the scale-invariant approximation, 
and all terms in Eqs. \eqref{eq:deltafNL2} 
are equally important.
Eqs. \eqref{eq:nmleading} and \eqref{eq:deltafNL1} are to be 
compared to Eq. (3.15) of Ref. \cite{Noller:2011hd}.
As pointed out in Ref. \cite{Burrage:2011hd}, 
there is no reason to believe these corrections will be negligible, 
but their magnitude will depend on the values of the parameters $\varepsilon$, 
$\eta$, $\delta$, and $n_s-1$. We do not attempt to produce an order
of magnitude of these corrections here, as this is beyond the scope of this paper.
We also notice that the dependence on the scale $k$ vanishes in the 
slow-roll limit, when $\varepsilon_\star$ is taken to be a perturbative 
parameter. This observation is in line with our previous comments on the 
strong scale-dependence whenever there was an appreciable deviation 
from the slow-roll regime, and the scale-invariant approach became appropriate.

\vspace*{0,3cm}

\para{Burrage \etal}
  Finally, our results extend what was obtained by 
  Burrage et al. \cite{Burrage:2011hd},
  and also Chen \etal \ \cite{Chen:2006nt} who, working in the 
  $P(X,\phi)$ class of models, 
treated
the speed of sound of scalar fluctuations, $s$, and the expansion rate, $\varepsilon$, 
as slowly varying. 
Taking the limit of small $\varepsilon$ and restoring $s$ 
in our formulae
by setting $\delta=s+2\varepsilon$, 
we find perfect agreement between our results to next-order
in slow-roll,
including the 
logarithmic corrections previously discussed.
As emphasized before, these corrections are important
to correctly evaluate the scale and shape-dependence
of the bispectrum, as well as to obtain an estimate 
of non-gaussianities with a certainty level comparable 
to the resolution of \textit{Planck}'s data.
As a consequence, our results obey Maldacena's factorization 
theorem \cite{Maldacena:2002vr, Creminelli:2004yq} in this limit, 
which represents a non trivial consistency check 
of our calculation.
The results presented  in Ref. \cite{Burrage:2011hd} 
regarding the scale and shape-dependence 
of the bispectrum apply to our analysis only if the 
slow-roll approximation is valid. Whenever there is significant 
breaking of the slow-roll approximation, and the scale-invariant
approximation applies, then a careful study of the scale and 
shape-dependence is required.

\section{Conclusions and outlook}
\label{sec:conclusions}

In the present era of precision Cosmology it is crucial 
to be able to use the experimental data 
soon to be delivered by \textit{Planck} efficiently. At the same time, 
the theoretical instruments at our disposal
should be able to compete with this level of accuracy. 
The most important of these are the search 
for non-gaussian signatures imprinted in the CMBR, 
which act as a model differentiator.
To this end, 
a primary object of study has been the bispectrum of primordial perturbations.

Whereas most studies of the bispectrum assume slow-roll
conditions, cosmological constraints still allow for a deviation from
this regime, whilst being compatible with
inflationary scenarios \cite{Khoury:2008wj}. In this paper we have focused on
the phenomenology of inflationary 
backgrounds where $\varepsilon$ and $s$ are \textit{not} perturbative 
parameters, and satisfy the mild requirement that 
$\varepsilon, |s|<1$. They are nevertheless combined 
to produce a scale-invariant spectrum 
of scalar perturbations. Under these assumptions we have calculated 
the bispectrum of single-field Horndeski models \textit{perturbatively}
in $n_s-1$, but to \textit{all orders} in $\varepsilon$ and $s$.
We have found that the scale-dependence 
of the bispectrum is encapsulated not only in a logarithmic \cite{Burrage:2011hd}, 
but in a stronger, power-law form. 
The power-law behaviour is more relevant if the breaking from
slow-roll is stronger, and the slow-roll approximation fails.

In an optimistic scenario, 
such behaviour can be used in our attempt to constrain
the parameters of a given theory more
tightly, and potentially 
eliminate it from the list of sensible models against observations. 
It is quite likely that with \textit{Planck} 
the amount of information one is able to 
extract from the CMBR about the early Universe will 
come to an end. It is therefore 
important to be able to retain the scale-dependence of the bispectrum
in our theoretical computations and estimations.
In particular, 
keeping the reference scale $k_\star$ arbitrary allows us to
 use these results for smaller scales 
(LSS) \cite{Sefusatti:2007ih, Afshordi:2008ru, Verde:2010wp, Desjacques:2010jw}.

In this paper we have focused on 
obtaining the general expression 
for the bispectrum in Horndeski theories
with fixed interaction coefficients, $\Lambda_i$. We have also
shown how to generalize our formulae to include rapidly varying $\Lambda_i$. 
This work should be regarded as a first step towards 
estimating observables on a background that 
does not obey the slow-roll regime. 
Because these theories have a strong scale-dependence, 
they are more vulnerable to data constraints arising from different scales.

We leave the study of decomposing bispectrum shapes
into fundamental harmonics (basis shapes)
for future work. This 
partial-wave decomposition is very important
in distinguishing between models, as discussed in Ref. \cite{Ribeiro:2011ax}. 
Also, it will be interesting to discuss the
subsequent implications in the consistency relations between the 
amplitudes of each harmonic, which can ultimately allow us to project out
this class of models based on a dynamics beyond slow-roll. 
If single field models are excluded by observations, we 
will have to embark on a multi-field
inflation exploration, where the interplay between curvature and 
isocurvature modes can be quite intricate
(see, for example, Refs. \cite{Byrnes:2010ft, Byrnes:2010xd, Seery:2012vj}) .

	\acknowledgments
I am grateful to David Seery
	for very helpful discussions, suggestions and encouragement in all the stages of this work.
	I thank Jo\~{a}o Magueijo, Johannes Noller, Gonzalo Palma 
	and Alexander Vikman for correspondence, 
and 	Daniel Baumann, Camille Bonvin and Ruth Durrer for discussions.  
	I also thank 
	Xingang Chen, 
	Anne Davis,  
	Matteo Fasiello,
	Federico Piazza,
	S\'{e}bastien Renaux-Petel, 
	and especially David Seery for useful comments on an early draft of this paper.
	I also appreciate the comments of the anonymous referee
	which have helped me to improve the first version of this paper.
	
	I have been supported by Funda\c{c}\~{a}o para a Ci\^{e}ncia e a Tecnologia 
	(Portugal) through the grant SFRH/BD/35984/2007, and 
	by the Cambridge Philosophical Society.
	\para{Note} This paper is dedicated to the memory of 
	Martinho Henriques and Cor\'{a}lia Ribeiro Norton.

\vspace*{1cm}

\begin{table}[htpb]

	\heavyrulewidth=.08em
	\lightrulewidth=.05em
	\cmidrulewidth=.03em
	\belowrulesep=.65ex
	\belowbottomsep=0pt
	\aboverulesep=.4ex
	\abovetopsep=0pt
	\cmidrulesep=\doublerulesep
	\cmidrulekern=.5em
	\defaultaddspace=.5em
	\renewcommand{\arraystretch}{1.6}
	\begin{center}
		\small
		\begin{tabular}{QqQqQ}

			\toprule
			\textrm{operator}
			&
			\multicolumn{4}{c}{contributions to $B^a$}
			\\
			\cmidrule(l){2-5}
		
		 	& 
		 	\multicolumn{1}{c}{$N^a(k_1)$} &
		 	\multicolumn{1}{c}{$P^a(k_1)$} &
		 	\multicolumn{1}{c}{$Q^a(k_1)$} &
		 	\multicolumn{1}{c}{$T^a(k_1)$}
		 	\\
			\midrule

			\zeta'^3 &
			12H_{\star} (1+\varepsilon_\star) \prod_i k_i^2 
			k_\star^{ \frac{5\varepsilon_\star}{1\varepsilon_\star}} & 
			-\im \bigg(\frac{1}{\im k_t} \bigg)^{3+ \frac{5\varepsilon_\star}{1+\varepsilon_\star}}
			&
			\\[2mm]

			\cmidrule{2-5}

			\zeta \zeta'^2 & 
			4 k_2^2 k_3^2 \,k_\star^{ \frac{4\varepsilon_\star}{1+\varepsilon_\star}} &
			\im \bigg(\frac{1}{\im k_t} \bigg)^{1+ \frac{4\varepsilon_\star}{1+\varepsilon_\star}}
			& 
			-k_1 \bigg(\frac{1}{\im k_t} \bigg)^{2+ \frac{4\varepsilon_\star}{1+\varepsilon_\star}}
			 &
			\\[2mm]

			\cmidrule{2-5}
	\multirow{3}*{$\zeta ( \partial \zeta)^2$} &
			\frac{4 \big(\vec{k}_2 \cdot \vec{k}_3 \big)}{c_{s\star}^2} &
			 &
			&
			\Omega_{1\star} \tilde{k}
			\\

			&
			 &
			&
			 &
			+\Omega_{2\star}\bigg(-k_t +\frac{k_1k_2k_3}{k_t^2}
			\bigg)			\\

			&
			 &
			&
			 &
			- \Omega_{2\star}\bigg(\EulerGamma- \ln\frac{k_\star}{k_t} \bigg) \tilde{k}
			\\[2mm]

			\cmidrule{2-5}
			
			\zeta' \partial_j \zeta \partial_j \partial^{-2} \zeta' &
			2k_1^2 \, (\vec{k}_2 \cdot \vec{k}_3)\, k_\star^{\frac{4\varepsilon_\star}{1+\varepsilon_\star}} &
			2\im\, \bigg(\frac{1}{\im k_t}\bigg)^{1+\frac{4\varepsilon_\star}{1+\varepsilon_\star}}  &
		-(k_2+k_3) \bigg(\frac{1}{\im k_t} \bigg)^{2+\frac{4\varepsilon_\star}{1+\varepsilon_\star}} \, & \\
[2mm]
			
			\cmidrule{2-5}
			
			\partial^2 \zeta (\partial_j \partial^{-2} \zeta')^2 &	 
			4k_1^2 \, (\vec{k}_2 \cdot \vec{k}_3)\, k_\star^{\frac{4\varepsilon_\star}{1+\varepsilon_\star}} &
			\im\, \bigg(\frac{1}{\im k_t}\bigg)^{1+\frac{4\varepsilon_\star}{1+\varepsilon_\star}} \, &
		-k_1\, \bigg(\frac{1}{\im k_t}\bigg)^{2+\frac{4\varepsilon_\star}{1+\varepsilon_\star}}  \,&		\\	
 			\bottomrule
	
		\end{tabular}
	\end{center}
	\caption{Coefficients of the functions appearing in the $B^a$ bispectrum.
			For simplicity of notation, we define the quantities
			$\tilde{k}$, $\Omega_1$ and $\Omega_2$ in table \ref{table:list}
			of appendix \ref{app:list}.
	\label{table:bafunctions}}
	\end{table}

\begin{table}[htpb]

	\heavyrulewidth=.08em
	\lightrulewidth=.05em
	\cmidrulewidth=.03em
	\belowrulesep=.65ex
	\belowbottomsep=0pt
	\aboverulesep=.4ex
	\abovetopsep=0pt
	\cmidrulesep=\doublerulesep
	\cmidrulekern=.5em
	\defaultaddspace=.5em
	\renewcommand{\arraystretch}{1.6}
	\begin{center}
		\small
		\begin{tabular}{QqQqQ}

			\toprule
			\textrm{operator}
			&
			\multicolumn{4}{c}{assignments to $\tilde{J}\gamma$ in $B^a$}
			\\
			\cmidrule(l){2-5}
		
		 	& 
		 	\multicolumn{1}{c}{$\gamma$} &
		 	\multicolumn{1}{c}{$A_\star$} &
		 	\multicolumn{1}{c}{$B_\star$} &
		 	\multicolumn{1}{c}{$C_\star$}
		 	\\
			\midrule

			\zeta'^3 &
			2+ \frac{5\varepsilon_\star}{1+\varepsilon_\star} & 
			\Omega_{3\star}&
			-\frac{h_{1\star}-5\beta_\star \varepsilon_\star}{1+\varepsilon_\star}
			+\frac{\delta_\star (2\varepsilon_\star -3)+\varepsilon_\star \eta_\star}{(1+\varepsilon_\star)^2}
		 &
			-\frac{5}{2} \frac{\alpha_\star \varepsilon_\star}{1+\varepsilon_\star}
			\\[2mm]

			\cmidrule{2-5}

			\zeta \zeta'^2 & 
			\frac{4\varepsilon_\star}{1+\varepsilon_\star} &
			\Omega_{1\star} & 
			-\frac{h_{2\star}-4\beta_\star \varepsilon_\star}{1+\varepsilon_\star}
			+\frac{\delta_\star (\varepsilon_\star -3)+2\varepsilon_\star \eta_\star}{(1+\varepsilon_\star)^2}
 &
						-2\frac{\alpha_\star \varepsilon_\star}{1+\varepsilon_\star}

			\\[2mm]
			\cmidrule{2-5}

			\zeta ( \partial \zeta)^2 &
			\multicolumn{4}{c}{not applicable}	\\[2mm]

			\cmidrule{2-5}
			
			\zeta' \partial_j \zeta \partial_j \partial^{-2} \zeta'  &
			\frac{4\varepsilon_\star}{1+\varepsilon_\star} &
			\Omega_{1\star} & 
			-\frac{h_{4\star}-4\beta_\star \varepsilon_\star}{1+\varepsilon_\star}
			+\frac{\delta_\star (\varepsilon_\star -3)+2\varepsilon_\star \eta_\star}{(1+\varepsilon_\star)^2} &
						-2 \frac{\alpha_\star \varepsilon_\star}{1+\varepsilon_\star} \\
		[2mm]
			
			\cmidrule{2-5}
			
			\partial^2 \zeta (\partial_j \partial^{-2} \zeta')^2 &	 
			\frac{4\varepsilon_\star}{1+\varepsilon_\star} &
			\Omega_{1\star} & 
				-\frac{h_{5\star}-4\beta_\star \varepsilon_\star}{1+\varepsilon_\star}
			+\frac{\delta_\star (\varepsilon_\star -3)+2\varepsilon_\star \eta_\star}{(1+\varepsilon_\star)^2}
			&
						-2 \frac{\alpha_\star \varepsilon_\star}{1+\varepsilon_\star} \\	
 			\bottomrule
	
		\end{tabular}
	\end{center}
	\caption{Coefficients of the functions $\tilde{J}_\gamma$ 
	appearing in the $B^a$ bispectrum, where
			$\Omega_{1\star}$ and $\Omega_{3\star}$
			are defined in table \ref{table:list} of appendix \ref{app:list}.
			The functions $\tilde{J}_\gamma$
			are defined in Eq. \eqref{eq:simplified_tildeJ}
			and discussed in detail in appendix \ref{app:other_integrals}.
	\label{table:JcoefficientsBa}}
	\end{table}

\begin{table}[htpb]

	\heavyrulewidth=.08em
	\lightrulewidth=.05em
	\cmidrulewidth=.03em
	\belowrulesep=.65ex
	\belowbottomsep=0pt
	\aboverulesep=.4ex
	\abovetopsep=0pt
	\cmidrulesep=\doublerulesep
	\cmidrulekern=.5em
	\defaultaddspace=.5em
	\renewcommand{\arraystretch}{1.6}
	\begin{center}
		\small
		\begin{tabular}{QqQqQq}

			\toprule
			\textrm{function}
			&
			\multicolumn{3}{c}{operator}
			\\
			\cmidrule(l){2-4}
		
		 	& 
		 	\multicolumn{1}{c}{$\zeta'^3$} &
		 	\multicolumn{1}{c}{$\zeta \zeta'^2$} &
		 	\multicolumn{1}{c}{$\zeta ( \partial \zeta)^2$} 
		 	\\
			\midrule

			N^b(k_1) &
			4H_{\star} (1+\varepsilon_\star) \prod_i k_i^2 
			k_\star^{ \frac{5\varepsilon_\star}{1+\varepsilon_\star}} & 
			4 k_2^2 k_3^2 \, k_\star^{ \frac{4\varepsilon_\star}{1+\varepsilon_\star}}
			& \frac{2 (\vec{k}_2 \cdot \vec{k}_3)}{c_{s\star}^2}  
			\\[2mm]

			\cmidrule{2-4}

P^b_1(k_1) & 
			\frac{1}{2}\bigg(\frac{1}{ \vartheta_1 k_t} \bigg)^{3+\frac{5\varepsilon_\star}{1+\varepsilon_\star}} \cos\bigg[\frac{5\pi \varepsilon_\star}{2(1+\varepsilon_\star)} \bigg] &
			\frac{1}{2}
			 \bigg(\frac{1}{ \vartheta_1 k_t} \bigg)^{1+\frac{4\varepsilon_\star}{1+\varepsilon_\star}}\cos\bigg[\frac{2\pi \varepsilon_\star}{1+\varepsilon_\star} \bigg]  & 
			-\frac{1}{2k_t} \big( k_1 k_2 +k_1 k_3- k_2 k_3\big)
			 
			\\[2mm]

			\cmidrule{2-4}
P^b_2(k_1)&
			\frac{1}{2}\bigg(\frac{1}{ \vartheta_2 k_t} \bigg)^{3+\frac{5\varepsilon_\star}{1+\varepsilon_\star}}\cos\bigg[\frac{5\pi \varepsilon_\star}{2(1+\varepsilon_\star)} \bigg]  &
		\frac{1}{2}
			\bigg(\frac{1}{ \vartheta_2 k_t} \bigg)^{1+\frac{4\varepsilon_\star}{1+\varepsilon_\star}} \cos\bigg[\frac{2\pi \varepsilon_\star}{1+\varepsilon_\star} \bigg] &
						-\frac{1}{2k_t} \big( k_1 k_2 +k_2 k_3- k_1 k_3\big)
		
			\\[2mm]

			\cmidrule{2-4}
			
P^b_3(k_1) &
			\frac{1}{2}\bigg(\frac{1}{ \vartheta_3 k_t} \bigg)^{3+\frac{5\varepsilon_\star}{1+\varepsilon_\star}} \cos\bigg[\frac{5\pi \varepsilon_\star}{2(1+\varepsilon_\star)} \bigg] &		
			\frac{1}{2}
			\bigg(\frac{1}{ \vartheta_3 k_t} \bigg)^{1+\frac{4\varepsilon_\star}{1+\varepsilon_\star}}
			\cos\bigg[\frac{2\pi \varepsilon_\star}{1+\varepsilon_\star} \bigg]  &
									-\frac{1}{2k_t} \big( k_1 k_3 +k_2 k_3- k_1 k_2\big)
\\
[2mm]
			
			\cmidrule{2-4}
			
Q^b_1(k_1) &	 
			 & 		-\frac{k_1}{2}
			\bigg(\frac{1}{ \vartheta_1 k_t} \bigg)^{2+\frac{4\varepsilon_\star}{1+\varepsilon_\star}}
			\cos\bigg[\frac{2\pi \varepsilon_\star}{1+\varepsilon_\star} \bigg] 
			 &
		-\frac{k_1k_2k_3}{k_t^2} \, \\	
[2mm]

			\cmidrule{2-4}
			
Q^b_2(k_1) &
			 &
			\frac{k_1}{2}
			\bigg(\frac{1}{ \vartheta_2 k_t} \bigg)^{2+\frac{4\varepsilon_\star}{1+\varepsilon_\star}} 
			\cos\bigg[\frac{2\pi \varepsilon_\star}{1+\varepsilon_\star} \bigg] 
			 & -\frac{k_1k_2k_3}{k_t^2} \\
[2mm]
			
			\cmidrule{2-4}
			
Q^b_3(k_1) &	 
		 &
			\frac{k_1}{2}
			\bigg(\frac{1}{ \vartheta_3 k_t} \bigg)^{2+\frac{4\varepsilon_\star}{1+\varepsilon_\star}}
			\cos\bigg[\frac{2\pi \varepsilon_\star}{1+\varepsilon_\star} \bigg]  &
		-\frac{k_1k_2k_3}{k_t^2}  \\	
		[2mm]

			\cmidrule{2-4}
R^b(k_1) &
			3\im \bigg( \frac{1}{\im kt}\bigg)^{3+\frac{5\varepsilon_\star}{1+\varepsilon_\star}}
 & 
			\im \bigg(\frac{1}{\im k_t} \bigg)^{1+ \frac{4\varepsilon_\star}{1+\varepsilon_\star}}
			
			& 
			\\[2mm]

			\cmidrule{2-4}

S^b(k_1) & 
			  &
			-k_1 \bigg(\frac{1}{\im k_t} \bigg)^{2+ \frac{4\varepsilon_\star}{1+\varepsilon_\star}}
			& 
		
			\\[2mm]

			\cmidrule{2-4}

		T^b(k_1) &
			
			  & -\bigg(\frac{1}{k_t}\bigg)^{1+\frac{4\varepsilon_\star}{1+\varepsilon_\star}} 
		\Xi
			 & f(k_1,k_2,k_3)

			\\
 			\bottomrule
 			
 					\end{tabular}
	\end{center}
	\caption{\label{table:bbfunctions1} Coefficients of the functions 
	appearing in the $B^b$ bispectrum for the first three operators, 
	where  $\vartheta_i=\frac{1}{k_t} (k_t-2 k_i) $,
	and $\Xi$ and
   $f(k_1,k_2,k_3)$ are defined in table \ref{table:list} of appendix \ref{app:list}.}
	
	\end{table}

\begin{table}[htpb]

	\heavyrulewidth=.08em
	\lightrulewidth=.05em
	\cmidrulewidth=.03em
	\belowrulesep=.65ex
	\belowbottomsep=0pt
	\aboverulesep=.4ex
	\abovetopsep=0pt
	\cmidrulesep=\doublerulesep
	\cmidrulekern=.5em
	\defaultaddspace=.5em
	\renewcommand{\arraystretch}{1.6}
	\begin{center}
		\small
		\begin{tabular}{QqQ}

			\toprule
			\textrm{function}
			&
			\multicolumn{2}{c}{operator}
			\\
			\cmidrule(l){2-3}
		
		 	& 
	
		 	\multicolumn{1}{c}{$\zeta' \partial_j \zeta \partial_j \partial^{-2} \zeta'$}&
		 	\multicolumn{1}{c}{$\partial^2 \zeta (\partial_j \partial^{-2} \zeta')^2 $}
		 	\\
			\midrule

			N^b &
		 2k_1^2 (\vec{k}_2\cdot \vec{k}_3) k_\star^{\frac{4\varepsilon_\star}{1+\varepsilon_\star}}
			&
					 4k_1^2 (\vec{k}_2\cdot \vec{k}_3) k_\star^{\frac{4\varepsilon_\star}{1+\varepsilon_\star}}
			\\[2mm]

			\cmidrule{2-3}

P^b_1 & 
\bigg(\frac{1}{\vartheta_1 k_t} \bigg)^{1+\frac{4\varepsilon_\star}{1+\varepsilon_\star}}\cos\bigg[ \frac{2\pi\varepsilon_\star}{1+\varepsilon_\star} \bigg] & 
\frac{1}{2} \bigg(\frac{1}{\vartheta_1 k_t} \bigg)^{1+\frac{4\varepsilon_\star}{1+\varepsilon_\star}}
\cos\bigg[ \frac{2\pi \varepsilon_\star}{1+\varepsilon_\star}\bigg]
			\\[2mm]

			\cmidrule{2-3}
P^b_2&
 \bigg(\frac{1}{\vartheta_2 k_t} \bigg)^{1+\frac{4\varepsilon_\star}{1+\varepsilon_\star}}\cos\bigg[ \frac{2\pi\varepsilon_\star}{1+\varepsilon_\star} \bigg] &
\frac{1}{2} \bigg(\frac{1}{\vartheta_2 k_t} \bigg)^{1+\frac{4\varepsilon_\star}{1+\varepsilon_\star}}
\cos\bigg[ \frac{2\pi \varepsilon_\star}{1+\varepsilon_\star}\bigg]			
			\\[2mm]

			\cmidrule{2-3}
			
P^b_3 &
		 \bigg(\frac{1}{\vartheta_3 k_t} \bigg)^{1+\frac{4\varepsilon_\star}{1+\varepsilon_\star}}\cos\bigg[ \frac{2\pi\varepsilon_\star}{1+\varepsilon_\star} \bigg]\, &
		 \frac{1}{2} \bigg(\frac{1}{\vartheta_3 k_t} \bigg)^{1+\frac{4\varepsilon_\star}{1+\varepsilon_\star}}
\cos\bigg[ \frac{2\pi \varepsilon_\star}{1+\varepsilon_\star}\bigg]
 \\
[2mm]
			
			\cmidrule{2-3}
			
Q^b_1 
		 &
		 \frac{k_2+k_3}{2} \bigg(\frac{1}{\vartheta_1 k_t} \bigg)^{2+\frac{4\varepsilon_\star}{1+\varepsilon_\star}}\cos\bigg[ \frac{2\pi\varepsilon_\star}{1+\varepsilon_\star} \bigg]
			& 
\frac{k_1}{2} \bigg(\frac{1}{\vartheta_1 k_t} \bigg)^{2+\frac{4\varepsilon_\star}{1+\varepsilon_\star}}
\cos\bigg[ \frac{2\pi \varepsilon_\star}{1+\varepsilon_\star}\bigg]			
			\\	
[2mm]

			\cmidrule{2-3}
			
Q^b_2 &

				\frac{k_3-k_2}{2}\bigg(\frac{1}{\vartheta_2 k_t} \bigg)^{2+\frac{4\varepsilon_\star}{1+\varepsilon_\star}}\cos\bigg[ \frac{2\pi\varepsilon_\star}{1+\varepsilon_\star} \bigg] &
				- \frac{k_1}{2} \bigg(\frac{1}{\vartheta_2 k_t} \bigg)^{2+\frac{4\varepsilon_\star}{1+\varepsilon_\star}}
\cos\bigg[ \frac{2\pi \varepsilon_\star}{1+\varepsilon_\star}\bigg]	
				 \\
[2mm]
			
			\cmidrule{2-3}
			
Q^b_3 &

			\frac{k_2-k_3}{2} \bigg(\frac{1}{\vartheta_3 k_t} \bigg)^{2+\frac{4\varepsilon_\star}{1+\varepsilon_\star}}\cos\bigg[ \frac{2\pi\varepsilon_\star}{1+\varepsilon_\star} \bigg] &
			-\frac{k_1}{2}\bigg(\frac{1}{\vartheta_3 k_t} \bigg)^{2+\frac{4\varepsilon_\star}{1+\varepsilon_\star}}
\cos\bigg[ \frac{2\pi \varepsilon_\star}{1+\varepsilon_\star}\bigg]	
						 \\	
		[2mm]

			\cmidrule{2-3}
R^b &

		-2 \im \bigg(\frac{1}{\im k_t} \bigg)^{1+\frac{4\varepsilon_\star}{1+\varepsilon_\star}}
			 &
			- \im\bigg(\frac{1}{\im k_t} \bigg)^{1+\frac{4\varepsilon_\star}{1+\varepsilon_\star}}
			\\[2mm]

			\cmidrule{2-3}

S^b & 

			  (k_2+k_3) \bigg(\frac{1}{\im k_t} \bigg)^{2+\frac{4\varepsilon_\star }{1+\varepsilon_\star}}
			&
			k_1 \bigg(\frac{1}{\im k_t} \bigg)^{2+\frac{4\varepsilon_\star}{1+\varepsilon_\star}}
			\\[2mm]

			\cmidrule{2-3}

			T^b &
			
			-2 \bigg(\frac{1}{ k_t} \bigg)^{1+\frac{4\varepsilon_\star}{1+\varepsilon_\star}}
			\Xi\,&

			-\, \bigg(\frac{1}{k_t} \bigg)^{1+\frac{4\varepsilon_\star}{1+\varepsilon_\star}}
						\Xi	

			\\
 			\bottomrule
 			
 					\end{tabular}
	\end{center}
	\caption{Coefficients of the functions appearing in the $B^b$ bispectrum
	for the last two operators, 
	where $\vartheta_i=\frac{1}{k_t} (k_t-2 k_i) $, and $\Xi$  
	is defined in table \ref{table:list} of appendix \ref{app:list}.
	\label{table:bbfunctions2}}
	\end{table}

\begin{table}[htpb]

	\heavyrulewidth=.08em
	\lightrulewidth=.05em
	\cmidrulewidth=.03em
	\belowrulesep=.65ex
	\belowbottomsep=0pt
	\aboverulesep=.4ex
	\abovetopsep=0pt
	\cmidrulesep=\doublerulesep
	\cmidrulekern=.5em
	\defaultaddspace=.5em
	\renewcommand{\arraystretch}{1.6}
	\begin{center}
		\small
		\begin{tabular}{QqQqQ}

			\toprule
			\textrm{operator}
			&
			\multicolumn{4}{c}{assignments to $\tilde{J}\gamma$ in $B^b$}
			\\
			\cmidrule(l){2-5}
		
		 	& 
		 	\multicolumn{1}{c}{$\gamma$} &
		 	\multicolumn{1}{c}{$A_\star$} &
		 	\multicolumn{1}{c}{$B_\star$} &
		 	\multicolumn{1}{c}{$C_\star$}
		 	\\
			\midrule

			\zeta'^3 &
			2+ \frac{5\varepsilon_\star}{1+\varepsilon_\star} & 
			\frac{\delta_\star}{1+\varepsilon_\star}- \frac{\varepsilon_\star\eta_\star}{(1+\varepsilon_\star)^2}+\frac{ \im  \pi}{4} (n_s-1) &
	\frac{n_s-1}{2} &

			\\[2mm]

			\cmidrule{2-5}

			\zeta \zeta'^2 & 
			\frac{4\varepsilon_\star}{1+\varepsilon_\star} &
		-\frac{3\delta_\star}{1+\varepsilon_\star}+ \frac{3\varepsilon_\star\eta_\star}{(1+\varepsilon_\star)^2}-\frac{3\pi \im }{4} (n_s-1) & 
			-\frac{3}{2} (n_s-1)
	&

			\\[2mm]
			\cmidrule{2-5}

			\zeta ( \partial \zeta)^2 &
			0 &\multicolumn{3}{c}{not applicable}	\\[2mm]

			\cmidrule{2-5}
			
			\zeta' \partial_j \zeta \partial_j \partial^{-2} \zeta'  &
			\frac{4\varepsilon_\star}{1+\varepsilon_\star}&
		\frac{3\delta_\star}{1+\varepsilon_\star} -\frac{3\varepsilon_\star \eta_\star}{(1+\varepsilon_\star)^2}+\frac{3\pi \im}{4 }(n_s-1)  &
			\frac{3}{2} (n_s-1) 
		 & 
					\\
		[2mm]
			
			\cmidrule{2-5}
			
			\partial^2 \zeta (\partial_j \partial^{-2} \zeta')^2 &	 
			\frac{4\varepsilon_\star}{1+\varepsilon_\star} &
				\frac{3\delta_\star}{1+\varepsilon_\star} -\frac{3\varepsilon_\star \eta_\star}{(1+\varepsilon_\star)^2}+ \frac{3\pi \im}{4 }(n_s-1)
			 & \frac{3}{2} (n_s-1)
 & 
					\\	
 			\bottomrule
	
		\end{tabular}
	\end{center}
	\caption{Coefficients of $\tilde{J}_\gamma$ appearing in the $B^b$ bispectrum.
		The functions $\tilde{J}_\gamma$
			are defined in Eq. \eqref{eq:simplified_tildeJ}
			and discussed in detail in appendix \ref{app:other_integrals}.
	\label{table:IJcoefficientsBb}}
	\end{table}

\newpage
\clearpage

\appendix
\section*{\begin{Large}
Appendices
\end{Large}}

\section{Dynamics in $y$---why?}
\label{appendix:y}
In this appendix we motivate the use of the time coordinate $y$,
instead of conformal time, $\tau$.
Why is this choice of space-time foliation, 
which defines a hypersurface at each $y=\textrm{constant}$, 
preferred compared to 
the more traditional conformal time? 
Were we to solve for the perturbations starting from 
Eq. \eqref{eq:quadratic action1} in $\tau$ coordinates, we would
have
obtained a formula for the propagator for scalar perturbations, 
$G_k$, which corresponds to 
the two-point correlator:
\begin{equation}
\langle \zeta (\vec{x},\tau) \zeta (\vec{y},\tilde{\tau}) \rangle =
G(\vec{x}, \tau; \vec{y},\tilde{\tau})\ \ .
\end{equation}
We would proceed to solve the Green's function equation for the propagator, 
in Fourier space, which reads
\begin{equation}
\bigg\{
\dfrac{\d ^2}{\d \tau^2}+
\bigg(\dfrac{1}{z}\dfrac{\d z}{\d\tau}+ \dfrac{2}{a}\dfrac{\d a}{\d\tau}  \bigg)
 \dfrac{\d}{\d\tau} 
+k^2 c_s^2
\bigg\} G_k (\tau ; \tilde{\tau}) = -\dfrac{\im}{2a^2 z} \delta (\tau-\tilde{\tau})\ \ ,
\label{eq:appb2}
\end{equation}
where $k$ is the comoving wavenumber and the Dirac delta
enforces evaluation at $\tau=\tilde{\tau}$.
Because scalar fluctuations propagate at a phase 
velocity $c_s$, generically different from that of the light, 
one usually performs a change of variables, $z=-k c_s \tau$, so that 
the propagator is a function of the sound horizon.
To get the equation for the evolution of $G_k$ in $z$ would demand inverting
$\d z/\d \tau$. When plugging into the equations of motion,
without the premiss of working 
with perturbative $s$, this would become algebraically challenging 
since we would be unable to truncate the Taylor expansion.
In particular, it would be very hard to show that the propagator 
is explicitly symmetric under the interchange $\tau \leftrightarrow \tilde{\tau}$.
It turns out that expressing the $y$ evolution of background quantities 
allows to naturally 
accommodate a rapidly varying speed of sound for the scalar perturbations, 
and therefore large $s$,
avoiding the difficulty just described.
In other words, the $y$ variable allows to sum all the powers 
in $s$ in Eq. \eqref{eq:appb2}. 

Moreover, writing the quadratic action for the fluctuations in 
the form \eqref{eq:quadratic action2} makes the reproduction of 
Bessel's equation more transparent, without any need for perturbative expansions. 
For these reasons, our dynamical analysis is presented in $y$-time, and 
follows the same lines of analysis by Khoury \& Piazza in Ref. \cite{Khoury:2008wj}.

\section{The spectral index beyond exact scale-invariance}
\label{app:ns}
In our main formulae we use $n_s-1$ as a perturbative parameter, but 
never its explicit formula, since it is never necessary
throughout the calculation. We have surpassed the need to know this
because of the special properties of the two-point correlator of 
single-field models, in particular that it should be 
time-independent on super-horizon scales.
  In this short appendix we 
present the formula for $n_s-1$ for completeness.
In the action \eqref{eq:quadratic action2} the variable $q=a\sqrt{z c_s}$
obeys the following differential equation
\[ \dfrac{\d \ln q}{\d\ln y} =-\dfrac{1}{1-\varepsilon -s}
\bigg\{1+\dfrac{w}{2} +\dfrac{s}{2} \bigg\}  \, 
\bigg\{1+\dfrac{\varepsilon \eta +t s}{(1-\varepsilon -s)^2} 
\bigg\} \ \ .\]
From this it follows that 
\begin{equation}
\dfrac{q''}{q} = \dfrac{1}{2y^2} \dfrac{\rho}{1-\varepsilon -s} +
\dfrac{1}{2y^2}\dfrac{1}{(1-\varepsilon -s)^2}
\bigg\{
\frac{\rho^2}{2}+wx+ts+
\dfrac{\varepsilon \eta +t s}{(1-\varepsilon -s)^2}
\rho  (4-2\varepsilon -s+w )
\bigg\} \ \ ,
\end{equation}
where $\rho\equiv 2+w+s$ and $x\equiv\d \ln w/\d N$ 
(which contributes at next-order only 
in the scale-invariant approximation).
We conclude that the spectral index in these theories 
satisfies
\begin{equation}
n_s-1=3-\sqrt{1+\frac{2\rho}{1-\varepsilon -s} +\frac{2}{(1-\varepsilon -s)^2}
\bigg\{
\frac{\rho^2}{2}  +wx+ts+
\dfrac{\rho(\varepsilon \eta +t s)}{(1-\varepsilon -s)^2}
  (4-2\varepsilon -s+w )
\bigg\}  }\ \ .
\label{eq:ns}
\end{equation}
This result is valid up to next-order in the scale-invariant approximation.
To leading order in the scale-invariant approximation, 
we find
\begin{equation}
n_s-1=\dfrac{-2\varepsilon -3s-w}{1-\varepsilon -s}
\label{eq:qsi_relation}
\end{equation}
in agreement with Ref. \cite{Burrage:2011hd}.
If we instead assume that $\varepsilon$ and $s$ are 
strictly constant and focus only  on
$P(X,\phi)$ models, we recover the formula
deduced in Ref. \cite{Khoury:2008wj}  
\begin{equation}
n_s-1=- \dfrac{2\varepsilon+s}{1-\varepsilon -s}\ \ , 
\end{equation}
which implies the exact scale-invariance relation
$s=-2\varepsilon$. In the slow-roll approximation,
this reduces to
\begin{equation}
n_s-1 \simeq -2\varepsilon -s
\end{equation}
at leading order, 
which indeed agrees with the results from Ref. \cite{Burrage:2011hd}.

In the scale-invariant approximation scheme $\delta$ is a next-order quantity 
and the only requirement we impose is 
that $\delta=s+2\varepsilon$, in $P(X,\phi)$ models. 
The general formula for $\delta$ in all the Horndeski models will 
depend on the form of $w$, and therefore $x$. 
Using Eq. \eqref{eq:ns} we find
\begin{equation}
n_s-1= -\dfrac{\eta+\delta}{1+\varepsilon}
-\dfrac{2}{3(1+\varepsilon)^2}
\Big\{
2 \varepsilon x+ 4 \big( \varepsilon \eta +t s \big) - \varepsilon t
\Big\}\ \ .
\end{equation}
This formula agrees with our expectation that $\delta$ is indeed 
a parameter contributing at next-order only
in the scale-invariant approximation.

\section{Next-order corrections---useful formulae}
\label{app:propagator}
When applying the Schwinger-Keldysh formalism, we will summon
formulae for the corrections of the 
elementary wavefunctions for $\zeta$ at next-order in the 
scale-invariant approximation scheme. This is 
because one should perform a uniform expansion in the slow-variation parameters
which include the interaction vertices, but also the propagator for 
scalar perturbations. In this appendix we collect some of the
necessary formulae for this expansion.

\para{Background evolution}
As explained in the main text, we wish to 
study the evolution of the background quantities in the 
time coordinate $y$.
To do so, we make a Taylor expansion around the time $y$ of 
horizon crossing of some reference scale $k_\star$, and make 
a uniform expansion for small $\eta_\star$ and $t_\star$.
Such expansion is well defined provided we restrict our analysis to a few e-folds
after horizon crossing. This ensures that $\varepsilon$ and $s$
have not varied significantly up to that point, and 
we can treat $\eta$ and $t$ as perturbative parameters in the dynamics.
As justified in the main text, this is indeed a good approximation.

To get the $y$-evolution of the speed of sound, $c_s$, we start
by evaluating 
\begin{equation}
\dfrac{\d\ln c_s}{\d \ln(-k_\star y)}=-\dfrac{s_\star}{1-\varepsilon_\star-s_\star}
\bigg\{1+
 \dfrac{\varepsilon_\star\eta_\star+t_\star s_\star}{(1-\varepsilon_\star-s_\star)^2} 
-\Big[\dfrac{t_\star}{1-\varepsilon_\star - s_\star}+ \dfrac{\varepsilon_\star \eta_\star
+t_\star s_\star}{(1-\varepsilon_\star-s_\star)^2}
 \Big]
 \ln(-k_\star y) \bigg\}\ \ .
 \label{eq:logcs}
\end{equation}
To avoid cluttering the notation, we introduce
the following parameters which contribute 
at next-order only in the scale-invariant approximation:
	\begin{subequations}
		\begin{align}
			\alpha
			& = \dfrac{t }{1-\varepsilon  - s }+ \dfrac{\varepsilon  \eta
+t s}{(1-\varepsilon -s )^2}\ \ \textrm{and}\\ 
\beta  
			& =\dfrac{\varepsilon \eta +t  s}{(1-\varepsilon -s)^2} \ \ .
		\end{align}
			\label{eq:alphabeta}
	\end{subequations}
Integrating both sides of Eq. \eqref{eq:logcs}, we find
\begin{equation}
c_s=c_{s\star} (-k_\star y)^{-\frac{s_\star}{1-\varepsilon_\star -s_\star}}
\bigg\{1-\dfrac{s_\star}{1-\varepsilon_\star -s_\star}
\bigg[\beta_\star \ln (-k_\star y) -\dfrac{\alpha_\star}{2} \big(\ln(-k_\star y)\big)^2 \bigg]
 \bigg\}\ \ .
 \label{eq:cs}
\end{equation}
This formula simplifies if we write $s_\star=-2\varepsilon_\star +\delta_\star$
and work perturbatively in $\delta_\star$, therefore assuming
$\Or(\delta_\star)=\Or(\eta_\star)=\Or(t_\star)$. 
We obtain
\begin{equation}
c_s=c_{s\star} (-k_\star y)^{\frac{2\varepsilon_\star}{1+\varepsilon_\star}}
\bigg\{1+ \frac{2\varepsilon_\star}{1+\varepsilon_\star} 
\bigg[ 
\beta_\star +\dfrac{\delta_\star (\varepsilon_\star -1)}{2\varepsilon_\star (1+\varepsilon_\star)}
\bigg]\, \ln (-k_\star y)
-\frac{\alpha_\star \varepsilon_\star}{1+\varepsilon} \, \big(\ln (-k_\star y)\big)^2
 \bigg\}\ \ .
 \label{eq:cs2}
\end{equation}
In the limit when $\varepsilon$ and $s$ are both
strictly constant, thereby resulting in vanishing $\alpha_\star$ and $\beta_\star$, 
we reproduce the results of Ref. \cite{Khoury:2008wj}
\[ c_s \sim (-k_\star y)^{-\frac{s_\star}{1-\varepsilon_\star -s_\star}}\ \ , \]
where we have temporarily restored the dependence in $s$ through $\delta$.
The additional contributions appearing in Eq. \eqref{eq:cs2}
are precisely the corrections to this purely power-law behaviour
and are relevant whenever $\varepsilon$ and $s$ are slowly-varying.

Proceeding similarly to get the dynamical evolution of the Hubble parameter, 
we obtain
\begin{equation}
H=H_{\star} (-k_\star y)^{\frac{\varepsilon_\star}{1-\varepsilon_\star -s_\star}}
\bigg\{1+\dfrac{\varepsilon_\star}{1-\varepsilon_\star  -s_\star}
\bigg[\beta_\star \ln (-k_\star y) -\dfrac{\alpha_\star}{2} \big(\ln(-k_\star y)\big)^2 \bigg]
 \bigg\}\ \ , 
 \label{eq:h}
\end{equation}
which, in the limit of constant $\varepsilon$ and $s$, reduces to simply
\[ H \sim (-k_\star y)^{\frac{\varepsilon_\star}{1-\varepsilon_\star -s_\star}} \ \ .\]
Again, recasting this result in terms of only one non-perturbative 
parameter, we find
\begin{equation}
H=H_{\star} (-k_\star y)^{\frac{\varepsilon_\star}{1+\varepsilon_\star}}
\bigg\{1+\dfrac{\varepsilon_\star}{1+\varepsilon_\star }
\bigg[\bigg(\beta_\star +\frac{\delta_\star}{1+\varepsilon_\star} \bigg)\ln (-k_\star y) 
-\dfrac{\alpha_\star}{2} \big(\ln(-k_\star y)\big)^2 \bigg]
 \bigg\}\ \ .
 \label{eq:h2}
\end{equation}
The explicit formulae in Eqs. \eqref{eq:cs2} and \eqref{eq:h2}
are relevant when replacing in Eq. \eqref{eq:a} for the scale factor, $a(y)$.
They are valid up to next-order in the scale-invariant approximation.

\para{Derivatives of the elementary wavefunctions}
When calculating the bispectra of Horndeski operators which are at least once
$y$-differentiated, we will require the derivatives of the 
elementary wavefunctions. The evolution of the background
primordial perturbation is given by
\begin{equation}
\dfrac{\d }{\d y}\zeta_k^{\textrm{(background)}} =
\dfrac{\im H_\star (1+\varepsilon_\star)}{2\sqrt{z_\star}\, (k\, c_{s\star})^{3/2}}
\, k^2 \, y \, e^{\im ky}\ \ ,
\end{equation}
whereas the wavefunction corrections to the internal lines obey
\begin{equation}
		\begin{split}
\dfrac{\d }{\d y}\delta \zeta_k^{\textrm{(internal)}}=
\dfrac{\im H_\star (1+\varepsilon_\star)}{2\sqrt{z_\star}\, (k c_{s\star})^{3/2}}
k^2 (-y)
\bigg\{ &
-\dfrac{n_s-1}{2}  e^{-\im ky} \int_{-\infty}^{y}\dfrac{e^{2\im k\xi}}{\xi} \d\xi\\
&+e^{\im ky}
\bigg[
\dfrac{\delta_\star}{1+\varepsilon_\star } -
\dfrac{\varepsilon_\star \eta_\star}{(1+\varepsilon_\star)^2} 
+\im \dfrac{\pi}{4} (n_s-1)
+\dfrac{n_s-1}{2}\ln (-k_\star y)
\bigg]
\bigg\} \ .
		\end{split}
		\label{eq:yderivatives}
\end{equation}
In obtaining these expressions we have explicitly eliminated the non-perturbative
dependence in the variation of the speed of sound, parametrized by $s$, 
in favour of the perturbative parameter, $\delta$.

The time variation of the wavefunctions is relevant 
in correctly reproducing the time-independence
of the correlation functions for $\zeta$, 
particularly in obtaining the corrections arising from 
the internal lines.

 \section{Useful integrals}
\label{app:integrals}
In this appendix we list the integrals which occur when computing the 
three-point correlators. They fall in essentially two different varieties:
integrals involving the exponential integral function 
(which first appeared in Eq. \eqref{eq:zetakinternal}),
and those which involve power 
laws and logarithmic integrand functions. We will identify master integrals 
for each of these families and analyse them separately.

\subsection{Integrals involving the exponential integral function, $\Ei(\xi)$}
\label{app:integral_function}
These integrals appear in the form
\begin{equation}
I_\gamma (k_3) \equiv 
\int_{-\infty}^0 {\d y}\, \bigg\{  {(-y)^\gamma}  
e^{\im (k_1+k_2-k_3) y}\, \int_{-\infty}^{y}{\dfrac{\d \xi}{\xi}\, e^{2\im k_3 \xi}} \bigg\} \ \ ,
\label{eq:integfunctionint}
\end{equation}
where we have explicitly written $I_\gamma$ as a function of the 
asymmetric momentum (in this case $k_3$), 
even though it is a function of the three
momenta through $k_t$ 
(perimeter in momentum space, $k_t=k_1+k_2+k_3$).
The constant $\gamma$ need not be an integer, 
and this is where the algebra differs
from that presented in Refs. \cite{Burrage:2010cu, Burrage:2011hd}. 
In what follows we consider positive, arbitrary $\gamma$.
To simplify Eq. \eqref{eq:integfunctionint}, one follows the algorithm
developed in these references: 
we apply a transformation of variables 
to convert this integral into its dimensionless 
version  by defining $x=\im (k_t-2k_3)y$).
We then make a rotation in the complex plane
using $w=-\im k_3 \xi$. 
Applying Cauchy's integral theorem we arrive at 
\begin{equation}
I_\gamma (k_3)= \bigg(\dfrac{1}{\im \vartheta_3 k_t}\bigg)
\ \int_{0}^{\infty}{\d x} \ \bigg\{ x^\gamma e^{-x} \ 
\int_{\infty}^{\theta_3 x} \dfrac{\d w}{w}\, e^{-2w}\bigg\}\ \ ,
\label{eq:integfunctionint2}
\end{equation}
where 
\[ \theta_3 \equiv\dfrac{k_3}{k_t-2k_3} \ \ \textrm{and}\ \ \ \vartheta_3\equiv 
\dfrac{1}{k_t} (k_t-2k_3) \ \ .\]
We make use of these abbreviated variables in \S \ref{subsec:bispectrum}.
Focusing on the inner integral in Eq. \eqref{eq:integfunctionint2}, we note that
upon integration by parts and using the expansion series of the exponential 
function, it has a convergent series representation as follows
\begin{equation}
\int_\infty^{\theta_3 x}\dfrac{\d w}{w} \ e^{-2w}= \EulerGamma
+\ln(2\theta_3 x) +\sum_{n=1}^{+\infty}\dfrac{(-1)^n}{n!} \dfrac{(2\theta_3 x)^n}{n}\ \ .
\end{equation}
Plugging this result into Eq. \eqref{eq:integfunctionint}
gives
\begin{equation}
I_\gamma(k_3)=\bigg(\dfrac{1}{\im \vartheta_3 k_t} \bigg)^{\gamma+1}
\bigg\{
\Gamma(1+\gamma) \Big[ \EulerGamma+
\ln (2\theta_3) +\psi^{(0)}(1+\gamma)
\Big]+
\sum_{n=1}^{\infty}\dfrac{(-2\theta_3)^n (n+\gamma)!}{n\, n!}
\bigg\}\ \ ,
\label{eq:integfunctionint3}
\end{equation}
in which $\psi^{(0)}(z)$ denotes the polygamma function of order 
zero and argument $z$.\footnote{The polygamma function of order $m$, $\psi^{(m)} (z)$,
						is the $(m+1)^{\mathrm{th}}$ derivative of the
						logarithm of the gamma function, $\Gamma$.
						We will only require polygamma functions of order
						zero and one in our formulae.} 
In obtaining this result 
we used the fact that $\Re(\gamma)>-1$ for the first two contributions, 
whereas $\Re(\gamma)>-2$ for the last term. This will indeed be true for all
the integrals we analyse since $\gamma$ is strictly non-negative. 
Finally, we can perform the last sum 
in Eq. \eqref{eq:integfunctionint3} by observing 
that it converges for $\Re(\theta_3) \in \big]-1/2; 1/2 \big[$.
We conclude that 
the integral \eqref{eq:integfunctionint} has a closed form 
representation, given by
\begin{equation}
		\begin{split}
I_\gamma(k_3)=\bigg(\dfrac{1}{\im \vartheta_3 k_t} \bigg)^{\gamma+1} &
\bigg\{
\Gamma(1+\gamma) \big[
\EulerGamma +\ln(2\theta_3) +\psi^{(0)}(1+\gamma)
\big]\\
& -2\theta_3 \Gamma (2+\gamma) \ \ 
_3 F_2\Big( \{1,1,2+\gamma\}, \{2,2\}, -2\theta_3 \Big)
\bigg\}\ \ ,
\end{split}
\end{equation}
where $F$ is a (convergent) 
generalized hypergeometric function, 
$\mathrm{HypergeometricPFQ}$ \cite{mABR70a}. In the main text 
we use the following abbreviated notation 
\begin{equation}
I_\gamma(k_3)=\bigg(\dfrac{1}{\im \vartheta_3 k_t} \bigg)^{\gamma+1} 
\ \tilde{I}_\gamma (k_3)\ \ ,
\end{equation}
which, given that $\tilde{I}_\gamma$
is a real-valued function, 
makes the identification of the real part of the result
of the integral more transparent. Explicit results for 
positive integers values of $\gamma=0,1,2$ are listed in 
table \ref{table:tildeI_integrals}.

\vspace*{1cm}

	\begin{table}[h]

	\heavyrulewidth=.08em
	\lightrulewidth=.05em
	\cmidrulewidth=.03em
	\belowrulesep=.65ex
	\belowbottomsep=0pt
	\aboverulesep=.4ex
	\abovetopsep=0pt
	\cmidrulesep=\doublerulesep
	\cmidrulekern=.5em
	\defaultaddspace=.5em
	\renewcommand{\arraystretch}{1.6}

	\begin{center}
		\small
		\begin{tabular}{cc}

			\toprule
		
			$\tilde{I}_\gamma (k_3)$ & explicit result \\
			\midrule
		
			\rowcolor[gray]{0.9}
				$\tilde{I}_0(k_3)$ &
				$ \ln (1-\vartheta_3) $ 
				\\[2mm]

				$\tilde{I}_1(k_3)$ &
				$ \vartheta_3 + \ln (1-\vartheta_3) $ 
				\\[2mm]

			\rowcolor[gray]{0.9}
				$\tilde{I}_2 (k_3)$ &
				$
				 \vartheta_3 (2+\vartheta_3) +2 \ln (1-\vartheta_3) $   \\
 			\bottomrule
		\end{tabular}
	\end{center}
	\caption{Explicit results for $\tilde{I}_\gamma$ integrals
	with $\gamma=0,1,2$. As argued in Ref. \cite{Burrage:2011hd}
	the function $I_\gamma(k_3)$ has no singularities in the limit
	 when $k_2, k_3 \rightarrow 0$, 
	for which $\vartheta_3 \rightarrow 0$ 
	and, in fact, one finds $\tilde{I}_0\rightarrow - \vartheta_3$, 
	$\tilde{I}_1 \rightarrow (-1/2)\, \vartheta_3^2$
	and $\tilde{I}_2 \rightarrow (-2/3)\, \vartheta_3^3$, 
	for example. 
	These precise values guarantee
	that Maldacena's consistency condition \cite{Maldacena:2002vr, Creminelli:2004yq}, 
	which relates the bispectrum evaluated 
	in the squeezed limit to the spectral index, is obeyed.
	\label{table:tildeI_integrals}}
	\end{table}

\subsection{Other integrals}
\label{app:other_integrals}

The other family of integrals which 
arises in the calculation of the bispectrum is of the form
\begin{equation}
J_\gamma \equiv \int_{-\infty}^{0} \d y e^{\im k_t y} \ (-y)^\gamma 
\bigg\{
A_\star + B_\star \ln (-k_\star y)+C_\star \big(\ln (-k_\star y) \big)^2 
\bigg\}\ \ ,
\label{eq:Jintegral}
\end{equation}
where the constant coefficients $A_\star$, $B_\star$ and $C_\star$ contain, in 
general, leading and next-order contributions in the scale-invariant 
approximation. Provided $\Re(\gamma)>-1$, which is indeed true for all our 
integrals, \eqref{eq:Jintegral} converges and gives
\begin{equation}
\begin{split}
J_\gamma= \bigg(\dfrac{1}{\im k_t}\bigg)^{1+\gamma} \, \Gamma (1+\gamma) \
\bigg\{ &
A_\star + \bigg[\ln (k_\star/k_t)-\im \frac{\pi}{2} \bigg]\, 
\bigg[B_\star + C_\star \bigg(\ln (k_\star/k_t)-\im \frac{\pi}{2}  \bigg) \bigg]+\\
& +\psi^{(0)} (1+\gamma) \bigg[
B_\star +2C_\star \bigg(\ln(k_\star/k_t)-\im \frac{\pi}{2} \bigg)
+C_\star \psi^{(0)} (1+\gamma)
\bigg]+ \\
&+ C_\star \psi^{(1)} (1+\gamma)
\bigg\} \ \ .
\end{split}
\label{eq:this_integral}
\end{equation}
Again, to simplify the notation we will refer to this integral in the form
\begin{equation}
J_\gamma \equiv \bigg(\dfrac{1}{\im k_t}\bigg)^{1+\gamma} 
\ \tilde{J}_\gamma\ \ .
\label{eq:simplified_tildeJ}
\end{equation}
We note that $\tilde{J}_\gamma$ is a complex-valued function, but 
given we only require the real part of $J_\gamma$ 
we need to use Euler's function and write
\[ \bigg( \dfrac{1}{\im k_t} \bigg)^{1+\gamma} =
\bigg( \dfrac{1}{ k_t} \bigg)^{1+\gamma}  \, \bigg\{
\cos\Big[\frac{\pi}{2} (1+\gamma) \Big] -\im \sin \Big[\frac{\pi}{2} (1+\gamma) \Big]
\bigg\}
\ \ .\]
This explains the presence of trigonometric functions 
in tables \ref{table:bbfunctions1} and \ref{table:bbfunctions2}.
As a final remark, we note that these integrals involve 
logarithmic contributions of the form $\ln (k_\star/k_t)$, 
which are responsible for scale-dependence, 
precisely in the same way the power spectrum 
has a weak, logarithmic scale-dependence [cf. Eq. \eqref{eq:powerspectrum}].

\section{Listing variables}
\label{app:list}
In the main tables of \S \ref{subsec:bispectrum}, 
to simplify the formulae, we have introduced a compact 
notation for combinations of momenta and slow-variation parameters. 
This is summarised in the following table.

	\begin{table}[htpb]

	\heavyrulewidth=.08em
	\lightrulewidth=.05em
	\cmidrulewidth=.03em
	\belowrulesep=.65ex
	\belowbottomsep=0pt
	\aboverulesep=.4ex
	\abovetopsep=0pt
	\cmidrulesep=\doublerulesep
	\cmidrulekern=.5em
	\defaultaddspace=.5em
	\renewcommand{\arraystretch}{1.6}

	\begin{center}
		\small
		\begin{tabular}{cc}

			\toprule
		
			variable & mathematical expression \\
			\midrule
		
			\rowcolor[gray]{0.9}
				$\tilde{k}$ &
				$ -k_t +\frac{\kappa^2}{k_t}+\frac{k_1 k_2 k_3}{k_t^2} $ 
				\\[2mm]

				$\kappa^2$ &
				$k_1k_2+k_1k_3+k_2k_3 $ 
				\\[2mm]

			\rowcolor[gray]{0.9}
$\Omega_{1\star}$ &

				$1+3\tilde{E}_\star +\frac{n_s-1}{2} \ln \Big(\frac{k_1 k_2 k_3}{k_\star^3} \Big) +  \frac{2\delta_\star}{1+\varepsilon_\star}-\frac{2\varepsilon_\star \eta_\star}{(1+\varepsilon_\star)^2} $   \\
				 [2mm]

				$\Omega_{2\star}$ &
				$ - \frac{\delta_\star+h_{3\star}}{1+\varepsilon_\star}+
				\frac{2\varepsilon_\star \eta_\star}{(1+\varepsilon_\star)^2} $ 
				\\[2mm]

			\rowcolor[gray]{0.9}
				$\Omega_{3\star}$ &
				$
				 \Omega_{1\star}-\frac{\delta_\star}{1+\varepsilon_\star}+\frac{\varepsilon_\star \eta_\star}{(1+\varepsilon_\star)^2} $   \\
				  [2mm]

\multirow{3}*{$f(k_1,k_2,k_3)$} &
				$  (n_s-1)\big[-k_t \ln \frac{8k_1k_2k_3}{k_t^3} +2 \big(k_1 \ln2k_1/k_t
   +k_2 \ln2k_2/k_t  +  k_3 \ln2k_3/k_t \big)$ \\

   &  $+3\tilde{k} (\EulerGamma -\ln k_\star/k_t)  
   -3\frac{k_1k_2k_3}{k_t^2} +4k_t-2\kappa^2/k_t   \big]$\\

   &
   $-3\tilde{k} \Big(\frac{\delta_\star}{1+\varepsilon_\star}-\frac{\varepsilon_\star \eta_\star}{(1+\varepsilon_\star)^2} \Big) $ 

				\\   				  [2mm]

				   			\rowcolor[gray]{0.9}

					$\Xi$ &
				$  \Gamma\bigg(1+\frac{4\varepsilon_\star}{1+\varepsilon_\star} \bigg)
			  \cos\bigg[\frac{2\pi \varepsilon_\star}{1+\varepsilon_\star}\bigg] $ 
				\\
 			\bottomrule
		\end{tabular}
	\end{center}
	\caption{This table collects all the abbreviated variables used in tables
	\ref{table:bafunctions}--\ref{table:bbfunctions2}. 
	\label{table:list}}
	\end{table}

%
\newpage
	\bibliographystyle{JHEPmodplain}
	\bibliography{paper}

\end{document}